\documentclass[galaxies,article,pdftex,moreauthors]{Definitions/mdpi}
\usepackage{babel}
\usepackage{ulem}
\newcommand{\bdm}{\begin{displaymath}}
\newcommand{\edm}{\end{displaymath}}
\newcommand{\beq}{\begin{equation}}
\newcommand{\eeq}{\end{equation}}
\newcommand{\bfi}{\begin{figure}}
\newcommand{\efi}{\end{figure}}     
\newcommand{\btab}{\begin{table}}
\newcommand{\etab}{\end{table}}
\newcommand{\btabu}{\begin{tabular*}{15.5cm}}
\newcommand{\btabo}{\begin{tabular*}{7.25cm}}
\newcommand{\etabu}{\end{tabular*}}

\firstpage{1} 
\pubvolume{1}
\issuenum{1}
\articlenumber{0}
\pubyear{2022}
\copyrightyear{2022}
\datereceived{} 
\dateaccepted{} 
\datepublished{} 
\hreflink{https://doi.org/} 

\defcitealias{castor1975}{CAK}
\defcitealias{ppk1986}{PPK}
\defcitealias{kpp1987}{KPP}
\defcitealias{kppa1989}{KPPA}
\defcitealias{muller2008}{MV08}
\defcitealias{friend1986}{FA}
\defcitealias{villata1992}{V92}

\Title{Radiation-Driven Wind Hydrodynamics of Massive Stars: A Review}

\TitleCitation{Massive Stars Wind Hydrodynamics}

\Author{Michel Cur\'e$^{1}$\,\orcidA{} and  Ignacio Araya$^{2}$ \,\orcidB{}}

\AuthorNames{Michel Cur\'e and Ignacio Araya}

\AuthorCitation{Cur\'e \& Araya}

\address{$^{1}$ \quad Instituto de F\'isica y Astronom\'ia, Universidad de Valpara\'iso, Chile\\
$^{2}$ \quad Vicerrector\`ia de Investigaci\'on, Universidad Mayor, Chile}

\corres{Correspondence: michel.cure@uv.cl}

\abstract{Mass loss from massive stars plays a determining role in their evolution through the upper Hertzsprung-Russell diagram. The hydrodynamic theory that describes their steady-state winds is the line-driven wind theory (m-CAK). From this theory, the mass-loss rate and the velocity profile of the wind can be derived, and knowing these properly will have a profound impact on quantitative spectroscopy analysis from the spectra of these objects. 
Currently, the so-called $\beta$ law, which is an approximation for the fast solution, is widely used instead of the m-CAK hydrodynamics, and when the derived value is $\beta \gtrsim 1.2$, there is no hydrodynamic justification for these values.
This review focuses on  1) a detailed topological analysis of the equation of motion (EoM); 2) solving numerically the EoM for all three different (fast and two slow) wind solutions; 3) deriving analytical approximations for the velocity profile via the Lambert$W$ function and 4) give a discussion of the applicability of the slow solutions. }


\begin{document}
\nolinenumbers
\section{Introduction} 
At the beginning of the XX century, \citet{johnson1925,johnson1926} and \citet{milne1926} argued
that the force on ions in the atmosphere of a luminous star could be 
responsible for the ejection of these ions from the star. They also argued that the ejected ions should carry with them the corresponding number of electrons, 
strictly there should be no charge--current, but they did not realize at that time that the collisional coupling between ions and  protons would drag the rest of the plasma (mostly fully ionized Hydrogen),
with them as well, at least to supersonic velocities, and this theory was
laid aside. It was Chandrasekhar \cite{chandrasekhar1941,chandrasekhar1943}, who in the
context of globular cluster dynamics, developed the theory of collisions due to an inverse square law, and \citet{spitzer1956} applied Chandrasekhar's theory for collisions between charged particles.

\citet{morton1967} was the first to report far-ultraviolet observations of three OB supergiants from an Aerobee-sounding rocket. After this came {\it Copernicus}, the first satellite with a telescope on board, and since then it has been possible to obtain stellar spectra in the ultraviolet (UV) region. \citet{morton1967} found that the resonance lines of C IV, N V  and Si IV showed the  typical P-Cygni\footnote{See \citet{lamers1999}, section 2.2} profiles. He found that the displacements in the profiles of C IV $\lambda\lambda 1549.5 $ and 
Si IV $\lambda\lambda 1402.8 $ corresponding to outflow velocities in the range
1500--3000 km/s.

\citet{snowmorton1976} showed through a detailed survey that stars brighter than $M_{bol}\sim-6$ have strong P-Cygni profiles in their spectra and therefore \textit{lose  mass}. The same conclusion was arrived at by  \citet{abbott1982}, who compared the radiative force with the 
gravitational force and concluded that radiative forces could initialize and
maintain the mass-loss process for stars with an initial mass at the zero-age main sequence (ZAMS) of about 15 $M_\odot$ or greater. 

This mass-loss process (known as stellar wind), together with supernovae explosions, are the main 
contributors in supplying the interstellar medium (ISM) with nuclear-processed heavy elements and therefore influence not only the chemical evolution (and therefore star formation) but also
the energy equilibrium of the ISM and the Galaxy \citep[see][and references therein]{kudritzki2000,puls2008,vink2022}.

\citet{parker1958} was the first to develop the solar wind through a purely gas-dynamical theory, which was until the winds of massive stars were discovered, the only known stellar wind theory.
When this theory was applied to the winds of a 
typical O-star, the effective temperature necessary to reproduce the observed terminal velocities was of the order of $10^7$ K, a value that is completely excluded by the presence of lines such as Si IV, C IV and N V ions, which would be destroyed by collisional ionization at temperatures above $ 3 \times 10^5 $ K. It was, therefore, necessary to seek an alternative mechanism to drive the wind. The natural driven mechanism is the force due to the interaction of the radiation field on the wind plasma, and the simplest form is the force due to the continuum, i.e., the Thompson radiative acceleration. This force leads macroscopically to a decrease in the star's gravitational attraction by a constant factor (for O--stars between $0.3$ and $0.6$). It is then clear that the continuum force alone cannot produce a force that exceeds gravity and, therefore, cannot drive these kinds of winds.\\

\citet{lucy1970} resuscitated the proposal of Johnson and Milne and considered the force due to the absorption of spectral lines, but unlike the earlier authors, they considered the flow of the plasma as a whole rather than the selective ejection of specific ions. They calculated an upper limit on the force on the C IV line $\lambda\lambda1548$ finding that this exceeds the force of gravity by a factor of approximately a few hundred. Hydrostatic equilibrium in the outermost layers is not possible, and an outflow of material must occur. In their stellar wind model, Lucy and Solomon made a series of assumptions, for instance, that the wind is driven only by resonance lines. They found mass-loss rates for O-stars of two orders of magnitudes less than the values obtained from observations.\\

A significant step in the theory was made by Castor, Abbot \& Klein \cite{castor1975} (hereafter \citetalias{castor1975}), who realized that the force due to line absorption in a rapidly expanding envelope could be calculated using the Sobolev approximation \citep{sobolev1960,castor1974}. Then by developing a simple parameterization of the line force, using the point star approximation, they were able to construct an analytical wind model. Despite the number of approximations made in that work, e.g., they represented the line force by C III lines and calculated only  one  model for a typical O5 f star ($T_{\rm{eff}} = 49\,290$ K, $\log g  = 3.94\footnote{The surface gravity $g$ is given in CGS units, i.e., cm/s$^2$. The quantity $\log g$ is dimensionless, see \citet{unidades}}$ and $R/R_{\odot} = 13.8$) obtaining a mass-loss rate $\dot{M} = 6.61 \times 10^{-6} \, M_{\odot}/$year and a terminal velocity $ v_{\infty} = 1515\, $ km/s. The value of the mass-loss rate was of the same order of magnitude as the values obtained from observation, but the terminal velocity lay below the measured ones. They also gave analytical scaling relations for the mass-loss rates and terminal speeds as functions of the stellar parameters. These were widely used to prove (or disprove) the validity of the radiation-driven  (or line-driven) wind theory by comparison with the observations.\\

\citet{abbott1982} improved this theory by calculating the line force 
using a tabulation of ca. $250\,000$ lines, which was complete for the elements H to Zn in the ionization states I to VI. Currently, the non-local thermodynamic equilibrium code CMFGEN \citep{hillier2012} uses around $900\,000$ lines; or FASTWIND with 4 million lines \citet{pauldrach2001} \citep[see also][who used ca. 4 million lines]{lattimer2021}. Despite this immense effort to give a more realistic representation of the line force, evident discrepancies with the observations remained. Simultaneously and independently, \citet{friend1986} and \citet{ppk1986} (hereafter FA and PPK, respectively) calculated the influence of the finite cone angle correction on the dynamics of the wind \citep[described in the appendix from][]{castor1974}.
They found a much better agreement between the improved or \textit{modified} CAK theory (hereafter m-CAK) and the observations for the mass loss rate and the terminal velocity in a large domain in the Hertzsprung-Russell diagram.

The equation of motion of the m-CAK theory is a highly non-linear differential equation that has singular points, eigenvalues and solution branches \citep[see][]{castor1975,friend1986,ppk1986,bjorkman2005,cure2007,cure2011}.
Since it is challenging to solve this differential equation numerically, \citetalias{ppk1986} found that the velocity field, $v(r)$ from the m-CAK theory, can be described by a simple approximation, known as the $\beta$ law approximation (see below). In addition, \citet[][hereafter KPPA]{kppa1989} developed analytical approximations for the localization of the critical point, mass loss rate and terminal velocity with an agreement within 5\% for $v_{\infty}$ and 10\% for $\dot{M}$, when compared to the correct numerical calculations.

Radiation-driven stellar winds are hydrodynamic phenomena involving the flow of the outer layers of the atmospheres of massive stars. This review is focused on describing the investigation of the m-CAK hydrodynamic theory, its topology and its three known physical solutions.

Section \ref{chapradfor} presents the theory to calculate the radiation (line) force via an analytical description thanks to the Sobolev approximation. Section \ref{sec-HYD} introduces the m-CAK hydro\-dynamic theory, and its topological description is given in section \ref{sec-topo}. Section \ref{typesols} shows all three known physical solutions, whilst section \ref{analyticalsols} present analytical approximate solutions based on the Lambert$W$ function. Finally, in section \ref{DiscConclu}, we summarise the main topics of this review and discuss the applicability of slow solutions.

\section{The Radiation Force \label{chapradfor}}

 The exact calculation for the radiation force requires a knowledge of the radiation field (in all the lines and continua) and of the physical processes (scattering, absorption and emission) that contribute to the exchange of energy and momentum throughout the wind. 
 The radiation field is represented by the monochromatic specific intensity $I_\nu(\mu)$, 
$\mu$ is the cosine of the angle between the 
incoming beam and the velocity vector of the interacting particles.
Thus, the radiation force per unit of volume, at a distance r, exerted on a point particle per unit of time is equal to momentum removed from the incident radiation field ($\kappa \rho\,I(\mu)\,\mu/c$) integrated over all the scattering directions. This force is given by
\begin{equation}
\label{eq2-1}
{\bf{F}}^{\rm{rad}}(r) = 
\frac{4 \pi}{c}
\frac{1}{2}
\int_{0}^{\infty}
\int_{-1}^{1}
\: \kappa_{\nu}(r)\,\rho(r)
\, I_{\nu}  \, \mu \, d\!\mu \, d\!\nu,
\end{equation}

\noindent where the absorption coefficient $\kappa_\nu$ is given in units of cm$^{2}$\,g$^{-1}$. The net flux density comes from the interaction processes, integrated over the whole spectral range, between the radiation field emitted by the photosphere and the stellar wind of mass density $\rho$ at the distance $r$. 
Here, it is assumed that the emissivity (thermal emission and photon scattering) in
the expanding atmosphere is isotropic. Therefore, no net momentum change occurs from this process (see \cite{hubeny2015}, Chapter 20).


The absorption coefficient $\kappa_\nu$  consists of three main contributions:
\begin{equation}
\label{eq2-2}
\kappa_{\nu}  =
\kappa^{\rm{Th}} + 
\kappa_{\nu}^{\rm{cont}} + 
\kappa_{\nu}^{\rm{line}}, 
\end{equation}
where $\kappa^{\rm{Th}}$ represents the Thomson scattering,
$\kappa_{\nu}^{\rm{cont}}$ the contribution of bound-free and free-free
transitions and $\kappa_{\nu}^{\rm{line}}$ the sum of all line absorption 
coefficients at frequency
$\nu$.

The radiation force can be calculated by state-of-the-art non-local thermodynamic equilibrium (NLTE) radiative transfer codes such as  {\sc Fastwind} \cite{santolaya1997,puls2005},  {\sc Cmfgen} \cite{hillier1987,hillier1998,hillier2001,hillier2012} or  {\sc PoWR} \cite{hamann1987,todt2015}, but these calculations depend on the velocity and density profile used to describe the wind. 

\subsection{Radiative force due to electron scattering}
The interaction between photons and free electrons is described by a Compton
process (an excellent review of this process, including Monte-Carlo 
calculations can be found in \cite{pozdnyakov1983}). If photons with energy $h\nu \ll m_e c^2$ are scattered
by Maxwellian electrons\footnote{Electrons with a velocity distribution function given by the Maxwellian distribution} having $k T \ll m_e c^2$, the frequency shift
will be very small, but if the scattering process is repeated many times, 
the small amounts of energy exchanged between the electrons and photons can build 
up and give rise to substantial effects.

In the non-relativistic limit without the influence of quantum effects
($h \nu \ll m_e c^2$), and neglecting
the possible effects described above,
the scattering cross-section is frequency-independent 
and called the Thomson cross-section, namely:
\begin{equation}
\label{eq2-3}
\sigma^{\rm{Th}} =
\frac{8 \pi}{3} \frac{e^4}{m_e^2 c^4} .
\end{equation}

The value of this cross-section is $\sigma^{\rm{Th}} = 6.65 \times 10^{-25} \,\mathrm{cm^2}$
and the absorption coefficient is, therefore:
\begin{equation}
\label{eq2-4}
\kappa^{\rm{Th}} \rho\, = n_e\,\sigma^{\rm{Th}} .
\end{equation}

 Using this value ($\kappa^{\rm{Th}}$) in Eq. (\ref{eq2-2}) and integrating Eq. (\ref{eq2-1}), we obtain  the contribution of the Thomson scattering to the radiation force,
\begin{equation}
\label{eq2-5}
{\bf{F}}^{\rm{Th}} = 
n_e 
\frac{ \sigma^{\rm{Th}} \, L }{ 4 \pi c \, r^2} \, ,
\end{equation}
where $L$ is the luminosity of the star. The radiative acceleration on the electrons is then 
\begin{equation}
\label{eq2-6}
g_e^{\rm{Th}} = \frac{1}{m_e} \frac{ \sigma^{\rm{Th}} \, L }{ 4 \pi c r^2}.
\end{equation}

It is useful to define the ratio of the Thomson scattering force
and the gravitational force by:
\begin{equation}
\label{eq2-7}
\Gamma_e  = \frac{g_e^{\rm{Th}}}{g^{\rm{grav}}} = \frac{1}{m_e} \frac{ \sigma^{\rm{Th}}\, L }{ 4 \pi \,c \, G M_{*}}\, ,
\end{equation}
here $G$ is the gravitational constant and $M_{*}$ the star's mass.
In the standard one-component description of stellar winds, the force over 
the density of the plasma is given by:
\begin{equation}
\label{eq2-8}
g^{\rm{Th}} =  \left(\frac{n_e}{\rho} \right) \frac{ \sigma^{\rm{Th}} \,L }{ 4 \pi c \, r^2 }
\end{equation}
being $\rho = m_p n_p + \sum_{\rm{ions}} (m_i n_i) + n_e m_e $ the mass density. The principal contribution of the ions comes from Helium, and neglecting the electrons, $n_e m_e$, the density is
\begin{equation}
\rho \simeq m_p n_p (1 + A_{\rm{He}} Y_{\rm{He}})\, .
\end{equation} Here $A_{\rm{He}}$ is the atomic mass of a Helium atom,  $Y_{\rm{He}}$ is the relative abundance by number of Helium with respect to Hydrogen (the latter being described by the subscript $p$), and $m_p$ is the proton mass. Based on the conservation of charge, it is possible to express the electron number density as $n_e =n_p (1 + q_{\rm{He}} Y_{\rm{He}})$, where $q_{\rm{He}} = 0, 1, 2$ depending on the Helium ionisation state.

Thus, the ratio $n_e/\rho$ is:
\begin{equation}
\label{eq2-9}
\frac{n_e}{\rho} =
\frac{1}{m_p} \left( \frac{ 1 + q_{\rm{He}} Y_{\rm{He}} }{ 1 + A_{\rm{He}} Y_{\rm{He}} }  \right)
\end{equation}
and the acceleration is:
\begin{equation}
\label{eq2-10}
g^{\rm{Th}} =  \frac{1}{m_p} \left( \frac { 1 + q_{\rm{He}} Y_{\rm{He}} }{ 1 + A_{\rm{He}} Y_{\rm{He}} }  \right)\left( \frac { \sigma^{\rm{Th}} L  }{ 4 \pi c \, r^2 } 
\right)
\end{equation}
or
\begin{equation}
\label{eq2-11}
\Gamma_{e}  = \left( \frac{ 1 + q_{\rm{He}} Y_{\rm{He}}}{ 1 + A_{\rm{He}} Y_{\rm{He}}} \right)\left( \frac{ \sigma^{\rm{Th}}\, L }{ 4 \pi\, c \,m_p\, G M_{*} } 
\right)
\end{equation}

\noindent Quite often, the canonical value of $\kappa^{\rm Th} = 0.34$ cm$^{2}$~g$^{-1}$ is adopted, which
follows from assuming a fully ionised plasma at solar abundance. In addition, since the continuum of OB stars, near its maximum, is also optically thin in the lines, the contribution of the continuum to the total radiative force is neglected.

The next section provides a general description of the line force based on the Sobolev approximation (see, e.g., \citet{lamers1999} or \citet{hubeny2015}).

\subsection{Radiative force due to lines.}
 
The contribution to the radiation force due to the spectral lines in the wind of massive stars is provided by the momentum transfer of photons (via absorption and 
re-emission processes in optically thick lines) mainly from the most dominant ions (i.e., C, O, N, and the Fe-group). 
The proper calculation of the line force  (per unit volume) is given by:
\begin{equation}
\label{eq2-12}
F^{\rm{line}}(r) = \frac{2 \pi}{c} \, \sum_{l} \int_{0}^{\infty}\int_{-1}^{+1} \kappa_l(r)\,\rho(r)\,\phi_l(\nu,\mu,r) \, I_{\nu}(r,\mu)\, \mu \, d\mu \, d\nu .
\end{equation}

\noindent  where $\phi$ is the Gaussian absorption profile. The summation is over all the line transitions ($l$),  assuming non-overlapping lines, for which the wind is optically thick. $\kappa_l$ is the opacity coefficient (in cm$^{2}$\,g$^{-1}$) of lines formed between levels $l$ (lower) and $u$ (upper) with energy h\,$\nu_0$, 

\begin{equation}
   \kappa_l\,\rho\,= \frac{\pi\,e^2}{m_e\,c}\,f_l\,n_l\,\left(1-\frac{n_u\,g_l}{n_l\,g_u}\right).
\end{equation}

The number density $n_l$ and $n_u$ of ions in levels $l$ and $u$ are given in cm$^{-3}$, $g_l$ and $g_u$ are the corresponding statistical weights, and $f_l$ is the oscillator strength of the line.
The \citetalias{castor1975} theory allows us to find an analytical
expression for the line force in a moving media with large velocity gradients in terms of the macroscopic variables using the Sobolev approximation. However, this expression only applies to radiating flows in the non-relativistic regime.

\subsubsection{The Sobolev approximation}
In a moving plasma like the stellar wind, the interaction of radiation with matter 
can be better understood as follows.
Let's consider a single spectral line  thermally broadened 
 with a rest wavelength $\lambda_1$. A photon emitted from the stellar surface, with wavelength $\lambda_\ast < \lambda_1$, propagates without interacting with the matter until, due to the Doppler shift, it is scattered at the blue edge of the line in question.
Due to the expansion of the wind, the particles viewed from any direction from a certain position always appear to be receding. This means that independent of the scattered direction of the photon (forward or backwards), the distance travelled always causes its comoving wavelength to be red-shifted.\\ 
After many scatterings, the photon's wavelength has been shifted to the line's red edge,  and the interaction of this photon with the line ($\lambda_1$) ceases. The region in the wind where an incoming photon can interact with the ions
is called the interaction zone. It is also well known that the winds of massive stars reach terminal velocities of several times the sound speed, and the point at which the
wind velocity is equal to the sound speed (the sonic point) is very near to the photosphere. This means that almost all the region where stellar winds are found is supersonic.\\
This description corresponds to the Sobolev approximation \citep{sobolev1960}, where all the relevant physical quantities, such as the opacity, source function, etc., are considered constant in the interaction zone, i.e., the width of the interaction zone is small compared with a characteristic flow length. Thus, for a generic Doppler-broadened line profile, the Sobolev-length, $L_s$, is  defined as: 
\begin{equation}
\label{eq2-13}
L_s =  v_{\rm{th}} / (d v / d r) ,
\end{equation}
where $T_{\rm{eff}}$ is the star's effective temperature, $v_{th}=\sqrt{2\, k_B\, T_{\rm{eff}}/m_p}$  the thermal speed of the protons, $k_B$ the Boltzmann constant.

A characteristic length of the flow is
\begin{equation}
\label{eq2-13b}
L_c \simeq  v / (d v / d r)
\end{equation}

Typical values of thermal velocities in OB-types stars are about $7 - 20$ km/s
while terminal velocities are about $1000$-$3000$ km/s, (see, e.g., \citet{lamers1999,puls2008}). More recent measurements of terminal velocities based on observations performed in the frame of the ULLYSES collaboration \citep{ullyses2023} have been accomplished by \citet{Hawcroft2023}.

\subsubsection{The line-force due to a single-line}
\citet{castor1974} analysed in detail the Sobolev approximation in the context of stellar winds and showed that the force produced by the incoming radiation due to a single line can be expressed as\footnote{This equation is for the direct radiation force as no scattering contributions are included within the Sobolev approximation.}:
\begin{equation}
\label{eq2-14}
f^{\rm{line}} = \left( \frac{ F_\nu \,\Delta \nu_d } { c } \right) \left( \frac{ k_l } { \tau_l } \right) (1 - e^{-\tau_l}),
\end{equation}
where $\Delta \nu_d = v_{\rm{th}} \, \nu / c$ corresponds to the Doppler shift, $F_\nu$ is the flux of the radiation field at frequency $\nu$, $k_l$ is the monochromatic line absorption coefficient per unit mass, and 
\begin{equation}
\tau_l = \int \rho \, \phi(\nu,r) \, k_l \, d r
\end{equation}
is the optical depth.
Evaluating the optical depth for a normalized Gaussian profile and using the Sobolev approximation, we find:
\begin{equation}
\label{eq2-15}
\tau_l  = k_l \, \rho \, v_{\rm{th}} / (d v / d r) .
\end{equation}
\noindent With this expression, we can interpret the RHS of (\ref{eq2-14}) as:
\begin{itemize}
\item[i)] $(F_\nu \,\Delta \nu_d / c)$ is the rate of momentum emitted by the star
per unit area at frequency $\nu$ with bandwidth $\Delta \nu_d$,
\item[ii)] $(\tau_l/k_l) = \rho \, v_{\rm{th}} / (dv/dr) $ represents the amount of mass that can absorb this momentum, and 
\item[iii)] $(1 - e^{-\tau_l})$ is the probability that such an absorption occurs.\\
\end{itemize}

Then, by defining, 
\begin{equation}
\label{eq2-16}
t = \sigma_e \, \rho \, v_{\rm{th}} / (dv/dr) \, ,
\end{equation}
where $\sigma_e  = \sigma^{\rm{Th}} n_e / \rho$ corresponds to  the Thomson scattering absorption coefficient per density. In a moving medium, $t$ represents the optical depth that a line will have if its opacity is equal to its electron scattering opacity. Based on this definition, it is possible to  rewrite $t$ as
\begin{equation}
\label{eq2-17}
\tau_l  = \eta_l \, t 
\end{equation}
where $\eta_l = k_l/\sigma_e$. The first factor in (\ref{eq2-17}) is related only to line properties, and the second only to dynamic variables of the wind.

\subsubsection{The line-force due to a statistical distribution of line strength}


The total line force, due to the addition of all the single lines of the ions, for a point star approximation and  for non-overlapping single lines, is given by: 
\begin{equation}
\label{eq2-18}
f^{\rm{line}} = \sum_l \left({\frac{ F_\nu \Delta \nu_d}{c}} \right)_{\!l} \,\left( {\frac{ dv/dr }{\rho\, v_{\rm{th}}} } \right) \,(1 - e^{-\eta_l t}) .
\end{equation}

Expressing (\ref{eq2-18}) in terms of $\Delta \nu_d = \nu \, v_{\rm{th}} / c$, and
the relation, $F = L / 4 \pi r^2$, we obtain
\begin{equation}
\label{eq2-19}
f^{\rm{line}} =  \frac{L}{c^2} \left( \frac{ dv/dr }{4 \pi r^2} \right) \, \sum_l \left( {\frac{ L_\nu \, \nu}{L}} \right)_{\!l} (1 - e^{-\eta_l t})
\end{equation}

\citet{abbott1982} was the first to compile and publish  a list of ca. 250\,000 lines for atoms from H to Zn in ionisation stages I to VI. Based on such a line list \citep{hillier1999,noebauer2015,lattimer2021}, it is possible to derive a line strength distribution function \citep[][]{ppk1986,puls2000}. This distribution can be described as follows:
\begin{equation}
\label{eq2-20}
dN(k_l)  = \int_0^N \left( \frac{ L_\nu \, \nu}{L} \right)\, n(k_l,\nu) d \nu
\end{equation}
and represents the number of lines in the line-strength interval $(k_l, k_l + \Delta k_l)$ obtained from the total spectrum and weighted by the flux mean of line strength $(L_\nu \, \nu/L)$. Notice that in Eq. (\ref{eq2-20}) the distribution in frequency space of the lines is independent from the distribution in line strength. An alternative formulation of the line statistic is given by \citet{gayley1995} \citep[see also][]{lattimer2021}.

The logarithm of the number of lines can be fitted by a linear function, namely:
\begin{equation}
\label{eq2-21}
dN(k_l)  =
N_0\,
(1-\alpha)\,
{\left(k_l/\sigma_e \right)}^{\alpha - 2}\,
d\left(k_l/\sigma_e \right)
\end{equation}
being $N_0$ the number of lines (strong and weak) that effectively contribute to 
the line force. Typical values of the parameter $\alpha$ are 
$0.45 \leq \alpha \leq 0.7$ \citep{lamers1999,puls2000}. Notice that line force 
parameters 
are not free  but depend on the transfer problem in 
each individual star \citep[see][for a detailed description of the calculation of the  line-force parameters]{puls2000,noebauer2015,gormaz2019, lattimer2021, gormaz2021,poniatowski2022}.

Extending the sum in Eq. (\ref{eq2-19}) to an integral, we obtain the line force expression:
\begin{equation}
\label{eq2-22}
f^{\rm{line}} = \frac{L}{c^2} \, \frac{ (dv/dr) }{4 \pi r^2} \,N_0 \, (1 - \alpha) \, \int_{\sigma_e}^\infty (1 - e^{-\eta_l t}){\left(k_l/\sigma_e \right)}^{(\alpha - 2)} \, d\left(k_l/\sigma_e \right).
\end{equation}
Neglecting the lower limit of the integral, a valid approximation for stars of type OB, replacing it by zero and integrating, the line force becomes:
\begin{equation}
\label{eq2-23}
f^{\rm{line}} =  \frac{L}{c^2} \,\frac{ 1 }{4 \pi r^2} \,v_{\rm{th}} \,  \sigma_e \, N_0 \, (1 - \alpha) \, \Gamma \!(\alpha){\left(\frac{dv/dr} {\sigma_e \,\rho \,  v_{\rm{th}}} \right)}^{\alpha},
\end{equation}
where $\Gamma$ is the $\Gamma$-function. Then, dividing by the total density, we obtain the standard form of the line 
acceleration, 
\begin{equation}
\label{eq2-24}
g^{\rm{line}} =\frac {C}{r^2} {\left(r^2 \,v \,\frac{dv}{dr} \right)}^{\alpha} \, ,
\end{equation}
with
\begin{equation}
\label{eq2-25}
C  = \Gamma_{e} \, G M_{*} \, k {\left(\frac{4 \pi }{ \sigma_e \, v_{\rm{th}} \,\dot{M} } \right)}^{\alpha} \,,
\end{equation}
here, $\Gamma_{e}$ is the radiative acceleration due to Thomson scattering in terms of the gravitational acceleration, and $\dot{M}$ is the mass-loss rate. Here the continuity equation has been used, and the  variables, such as $N_0$ or $\Gamma(\alpha)$, have been collected into the constant $k$. Note that this expression for the line-force (Eq. \ref{eq2-24}) only takes  interactions between ions and radially emitted photons into account \citep{castor1974,castor1975}. 

\subsubsection{The correction factor}
 \citetalias{castor1975} (see their appendix) discussed qualitatively the effect on the line force that the proper shape of the star (non-radial incoming photons) would have on the wind kinematics. Later \citetalias{ppk1986} and \citetalias{friend1986} independently investigated the influence of this effect, known as the finite disk correction factor, thereby developing the m-CAK theory. 

The expression of the line force for incoming photons from an arbitrary direction, for a radial flow velocity field, comes from the definition of Eq.
(\ref{eq2-16}), thus,
\begin{equation}
\label{eq2-26}
t_{\sigma}=  \left(\frac{1+\sigma}{1 + \mu^2 \sigma} \right) \frac{ \sigma_e \, \rho \, v_{\rm{th}}}{dv/dr} ,
\end{equation}
where
\begin{equation}
\label{eq2-27}
\sigma = \frac{ d \ln v }{ d \ln r } .
\end{equation}

Inserting $t_{\sigma}$ in Eq. (\ref{eq2-22}), instead of $t$, and integrating, we 
obtain the following expression for the line force:
\begin{equation}
\label{eq2-28}
g^{\rm{line}} = \frac {C}{r^2} CF(r,v,v'){\left(r^2 \,v \,v' \right)}^{\alpha} 
\end{equation}
where $CF$  is the correction factor, defined as the ratio of the force due to the non-radial contributions to that of a point star approximation, namely:
\begin{equation}
\label{eq2-29}
CF = \frac {2}{1 - \mu_c^2} \int_{\mu_c}^{1} {\left[ \frac{ (1 - \mu^2) \, v/r + \mu^2 v^\prime}{ v^\prime } \right]}^{\alpha} \, \mu\, d\mu
\end{equation}
here $\mu_c = \sqrt{(1 - R_\ast^2/r^2)}$, where $R_{\ast}$ is the stellar radius,  and $v^\prime = dv/dr$. In  appendix \ref{appA}, we summarised some properties of the correction factor.

\subsection{The Ionization Balance}
In his work, \citet{abbott1982}, assumed local thermodynamic equilibrium (LTE) and used the modified Saha formula (see \citet[][]{hubeny2015}) to  take into account the dilution  of the radiation field and the possible difference between the  electron kinetic temperature $T_e$ and the radiation temperature $T_r$.
Due to the changes in the  ionisation throughout the wind, Abbott fitted the line force not only in terms of $(r^2\, v \,v')^\alpha$ (see Eq. \ref{eq2-24}) but also as a function of the ratio $n_e / W(r)$, where 
\begin{equation}
W(r) = \frac{1}{2} \left(1 - \sqrt{(1 - R_\ast^2/r^2)} \right) \, ,
\end{equation}
is the dilution factor. He found that the functional dependence of this quotient in the line force is:
\begin{equation}
\label{eq2-31}
g^{\rm{line}} \propto {\left(\frac{ n_e }{ W(r)} \right)}^\delta ,
\end{equation}
where the electron number density, $n_e$, is given in units of $10^{11} \rm{cm^{-3}}$.
This proportionality means that the greater the density, the lower the ionisation level. In view of the fact that the lower ionisation levels have more line transitions, usually at the maximum of the radiation field, the line force increases with increasing density. Values of this $\delta$ line-force parameters for the fast solution (see below) are in the range $0<\delta\lesssim 0.2$ \citep{lamers1999}, but for a pure Hydrogen atmosphere, the value is $\delta=1/3$ as \citet{puls2000} demonstrated.



\section{The m-CAK Hydrodynamic model \label{sec-HYD}} 
The 1-D m-CAK stationary model for line-driven stellar winds considers the following assumptions: an isothermal fluid in spherical symmetry and neglecting the influence of viscosity effects, heat conduction and magnetic fields.\\

The stationary continuity and momentum conservation equations are:  
\begin{equation} 
\dot{M} = 4\,\pi\, r^{2}\, \rho\, v \label{cont-eq} ,
\end{equation} 
\begin{equation} 
v\,\frac{dv}{dr}=-\frac{1}{\rho }\,\frac{{dP}}{dr}-\frac{GM_{*}(1-\Gamma_{e})}{r^{2}}+ 
\frac{v_{\phi }^{2}(r)}{r}+g^{\rm{line}}(\rho,dv/dr,n_{e})  \label{mom-eq} ,
\end{equation} 
being $P$ the fluid pressure, $v_{\phi }= v_{rot} R_{\ast} / r$, where $v_{rot}$ is the stellar rotational speed at the equator. In addition,   $g^{\rm{line}}(\rho,dv/dr,n_{e})$ corresponds to the acceleration due to an ensemble of lines.

The standard or m-CAK parameterization of the line force \cite{abbott1982,ppk1986,friend1986} is the following: 
\begin{equation} 
g^{\rm{line}}=\frac{C}{r^{2}}\;CF(r,v,dv/dr)\;\left( r^{2}v\frac{dv}{dr} 
\right) ^{\alpha }\;\left( \frac{n_{e}}{W(r)}\right) ^{\delta }  \label{2.2} , 
\end{equation} 
where the coefficient $C$ is given by Eq. (\ref{eq2-25}).

Substituting the density from the mass conservation equation (Eq.~\ref{cont-eq}) into the momentum equation (Eq.~\ref{mom-eq}), we obtain the equation of motion (EoM). 

Transforming to dimensionless variables, that is: 
\begin{eqnarray} 
u &=&-\frac{R_{\ast }}{r} \, \label{2.4a} ,\\ 
w &=& \frac{v}{a}  \,\label{2.4b} ,\\ 
w'&=& \frac{dw}{du} \, ,\label{2.4c}
\end{eqnarray}
where $a$ is the isothermal sound speed of an ideal gas, $P=a^{2}\rho$. \\

Using these new variables, the EoM now reads: 
\begin{equation} 
F(u,w,w') = \left( 1-\frac{1}{w^{2}} \right)
w\frac{dw}{du}+A+\frac{2}{u}+a_{\rm{rot}}^{2}u-C' 
\;CF\;g(u)(w)^{-\delta }\left( w\frac{dw}{du}\right) ^{\alpha 
}\ = 0 \, ,
\label{2.5} 
\end{equation} 
where the constants are the following: 
\begin{eqnarray} 
A&=&\frac{GM(1-\Gamma_{e} )}{a^{2}R_{\ast }}=\frac{v_{\rm{esc}}^{2}}{2a^{2}} \label{2.5d}, \\
a_{\rm{rot}}&=&\frac{v_{\rm{rot}}}{a}  \label{2.5e}, \\
C'&=&C\;\left( \frac{\dot{M} D}{2\pi}\frac{10^{-11}}{aR_{\ast
}^{2}} \right)^{\delta }\;(a^{2}R_{\ast })^{(\alpha -1)}  \label{2.5b}, \\
D&=& \frac{(1+Z_{\rm{He}} Y_{\rm{He})}}{(1+A_{\rm{He}} Y_{\rm{He}})}\frac{1}{m_{\rm{p}}} \label{2.5f},
\end{eqnarray}
being 
$Z_{\rm{He}}$ the number of free electrons provided by Helium,  $v_{\rm{esc}}$ the escape velocity, and 
the function $g$ is defined as: 
\begin{equation} 
g(u)=\left( \frac{u^{2}}{1-\sqrt{1-u^{2}}}\right)^{\delta }  \label{2.5c} 
\end{equation}

In order to find a physical wind solution of the EoM (Eq. \ref{2.5}), i.e., 
starting from the photosphere with a small velocity and reaching infinite with a supersonic velocity, we first need to understand the topology of this equation.

\section{Topological Analysis \label{sec-topo}}

As mentioned previously, the first wind model was developed by  \citet{parker1958} for the sun. This model possesses a
singular point at the sonic point and different solution branches \citep[see Fig. 3.1 from][]{lamers1999}. The m-CAK model has a driving force (line force) that depends not only on the radial coordinate $r$ (or $u$) but also on the velocity and the velocity gradient. These characteristics complicate the study of the EoM's topology that gives rise to the different solutions.\\

Mathematically, singular points are located where the \textit{singularity condition} is satisfied, i.e.:
\begin{equation} 
\frac{\partial}{\partial w'}F(u,w,w')=0  \label{2.6} \,,
\end{equation} 
and these locations form the \textit{locus} of singular points.

At these specific points, in order to have a smooth wind solution between solution branches, a \textit{regularity condition} must be imposed, namely:
\begin{equation} 
\frac{d}{du} F(u,w,w')=\frac{\partial F}{\partial u}+\frac{\partial 
F}{\partial w}w'=0  \label{2.7} 
\end{equation} 

Using the following coordinate transformation:  
\begin{equation} 
Y=w\;w'  \label{2.8a} 
\end{equation} 
\begin{equation} 
Z=w/w'  \label{2.8b} ,
\end{equation} 
we can now solve Equations (\ref{2.5}), (\ref{2.6}) and (\ref{2.7}), only valid simultaneously at one singular point, obtaining the following set of Equations:
\begin{eqnarray} 
Y -\frac{1}{Z}&+A+ 2/u + a_{rot }^{2}u & -\,C' 
f_{1}(u,Z)\,g(u)\,Z^{-\delta /2}\,Y^{\alpha -\delta /2} = 0 \; 
, \label{2.9a}\\ 
Y -\frac{1}{Z} & & -\,C' 
f_{2}(u,Z)\,g(u)\,Z^{-\delta /2}\,Y^{\alpha -\delta /2} = 0 \; , \label{2.9b}\\ 
Y +\frac{1}{Z} &-2Z/u^{2} +a_{rot}^{2}Z\ & -\,C' 
f_{3}(u,Z)\,g(u)\,Z^{-\delta /2}\,Y^{\alpha -\delta /2} = 0 \; , \label{2.9c}
\end{eqnarray}
derivation details and definitions of $f_{1}(u,Z)$,
$f_{2}(u,Z)$ and $f_{3}(u,Z)$ are summarised in appendix~\ref{appB}.

Solving for $Y$ and $C'$ from the set of Equations (\ref{2.9a}),
(\ref{2.9b}) and (\ref{2.9c}), we obtain:  
\begin{equation} 
Y=\frac{1}{Z}+\left( \frac{f_{2}(u,Z)}{f_{1}(u,Z)-f_{2}(u,Z)}\right) \times \left( 
A+\frac{2}{u}+a_{rot}^{2}u\right)   \label{2.10} 
\end{equation} 
and 
\begin{equation} 
C'(\dot{M})=\frac{1}{g f_{2}}\left( 1-\frac{1}{Y\,Z}\right) \;Z^{\delta 
/2}\;Y^{1-\alpha +\delta /2}  \label{2.11} 
\end{equation} 
These  last two Eqs. are generalisations of the relations found by \citetalias{kppa1989} (see
their Eq.[21] and Eq.[34] for $Y$ and Eq.[20] and Eq.[44] for the eigenvalue), but  now including the rotational speed of the star.

\subsection{The critical point function $R$  \label{sec-RUZ}}

The set of Eqs. (\ref{2.9a}), (\ref{2.9b}) and (\ref{2.9c}) are only 
valid at the singular point for the unknowns: $Y_s$, $C'_s$, $Z_s$ and $u_s$\footnote{The subscript \textit{s} means at the singular point.}.
Due to the fact that there are only three equations and four unknowns, it is not possible to solve them. Nevertheless, from this set of Equations, we can derive the function, $R(u,Z)$, defined by:
\begin{equation} 
R(u,Z)= -\frac{2}{Z} + \frac{2Z}{u^{2}} - a_{\rm{rot}}^{2}Z
+ f_{123}(u,Z)\left( A+ \frac{2}{u}+a_{\rm{rot}}^{2}u\right)   
\label{2.12} 
\end{equation} 
where $f_{123}(u,Z)$ has the following definition:
\begin{equation} 
f_{123}(u,Z) = \frac{f_{2}(u,Z)-f_{3}(u,Z)}{f_{2}(u,Z)-f_{1}(u,Z)}
 \label{2.12b} 
\end{equation}
The locus of singular points, $u_{s}$, is given by the  points
which are solutions of the following equation:
\begin{equation} 
R(u,Z) = 0 \,.
\label{2.12b2} 
\end{equation}
It should be noted that no approximation has been made in the derivation of the above topological equations.

To determine which is the location of the singular point in the locus of points that satisfies,
$R(u,Z) = 0,$ and therefore determine the values of $Y_s$,$C'_s$, $Z_s$ and $u_s$, we need to
set a boundary condition at the stellar surface.\\
For this boundary condition, the most used are:
\begin{itemize}
\item[i)] Set the density at the stellar surface to a specific value,
\begin{equation} 
\rho(R_{*}) =  \rho_{*} ,
\label{2.5ab} 
\end{equation} 
Usually this base density is in the range $10^{-8}\, \rm{g\,cm^{-3}}$ to $10^{-13}\, \rm{g\,cm^{-3}}$. For some examples, see the works of \citet{araujo1989,friend1984, madura2007, cure2004, araya2018}.

\item[ii)] Set the optical depth integral to a specific value, i.e.,
\begin{equation} 
\tau_{*}=\int_{R_{\ast}}^{\infty} \sigma_{E} \, \rho(r) dr = \frac{2}{3} .
  \label{2.5a} 
\end{equation} 
\end{itemize}

Employing one of these boundary conditions at the stellar surface plus the regularity condition at the singular point, we can solve from the EoM (Eq. \ref{2.5}) the velocity profile, $w=w(u)$, together with the value of the eigenvalue, $C'$, and therefore the mass loss rate, $\dot{M}$.

\section{Types of solutions \label{typesols}}
We developed a numerical code that discretizes by finite differences the EoM and, using the Newton-Raphson method, iterates to a numerical solution, this code is called {\sc Hydwind} and is described in more detail in \citet{cure2004} 
\citep[see also ][]{cure1992}.\\

After performing the topological analysis of the EoM, we were able, thanks to {\sc Hydwind}, to find the numerical solutions of all three known m-CAK physical solutions: fast, $\Omega$-slow and $\delta$-slow solutions.

\subsection{Fast solution \label{fastsol}}

From the pioneering work of \citetalias{castor1975} and its improvements from \citetalias{friend1986} and  \citetalias{ppk1986}, the code {\sc Hydwind} is able to obtain the standard solution of the m-CAK theory, and we called it hereafter the \textit{fast} solution. 

To perform our topological analysis, we  use a typical O5 V star with the following stellar parameters, $T_{\rm eff}=45000K$, $\log g=4.0$, $R/R_{\odot}=12$, and line force parameters $k=0.124$, $\alpha=0.64$, and $\delta=0.07$, with the boundary condition  $\tau_{*}=2/3$.

\begin{figure}[H]
\center
\includegraphics[width=6.8cm]{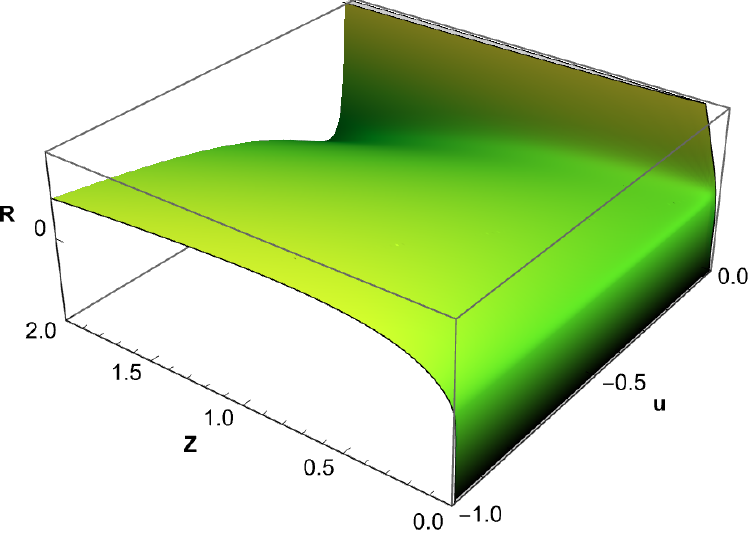}
\includegraphics[width=6.8cm]{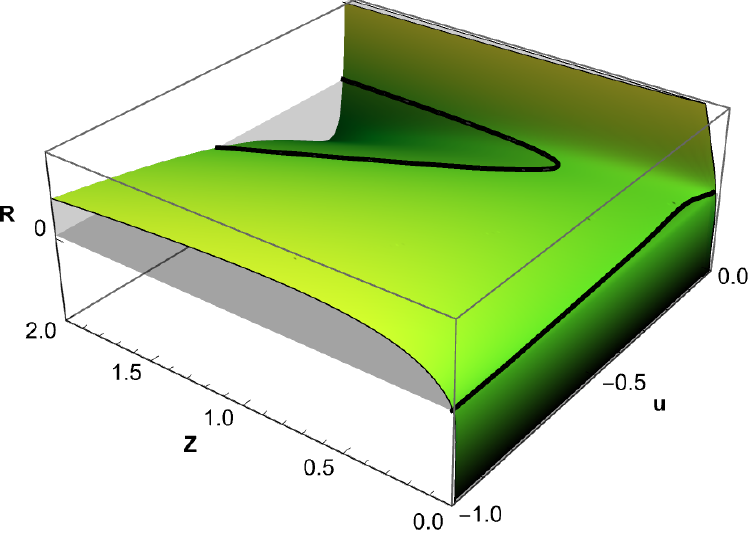}
\caption{Function $R(u,Z)$ for a typical O5 V star without rotation. The left panel shows only the function $R(u,Z)$, while the right panel is similar to the left panel, but the plane $R(u,Z)=0$ is also plotted in light grey. Furthermore, the intersection of both curves (black solid lines) shows two locus of singular points.
\label{fig1}}
\end{figure}

The function $R(u,Z)$ is shown in Fig. \ref{fig1} for a non-rotating star ($a_{\rm{rot}}=0$). The plane $R(u,Z)=0$ is plotted in light grey (right panel) and the intersection of both functions, which corresponds to the locus of singular points, is plotted with black lines.  The locus of singular points for the fast solution is the one that starts at $Z\sim0$ and $u=-1$. The other locus of singular points will be discussed below.

\begin{figure}[H]
\center
\includegraphics[width=6.8cm]{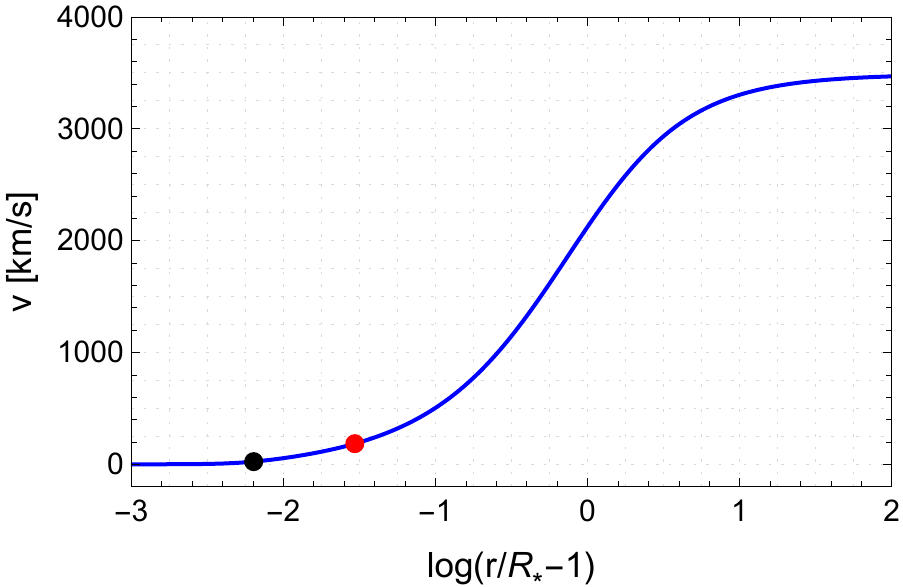}
\includegraphics[width=6.8cm]{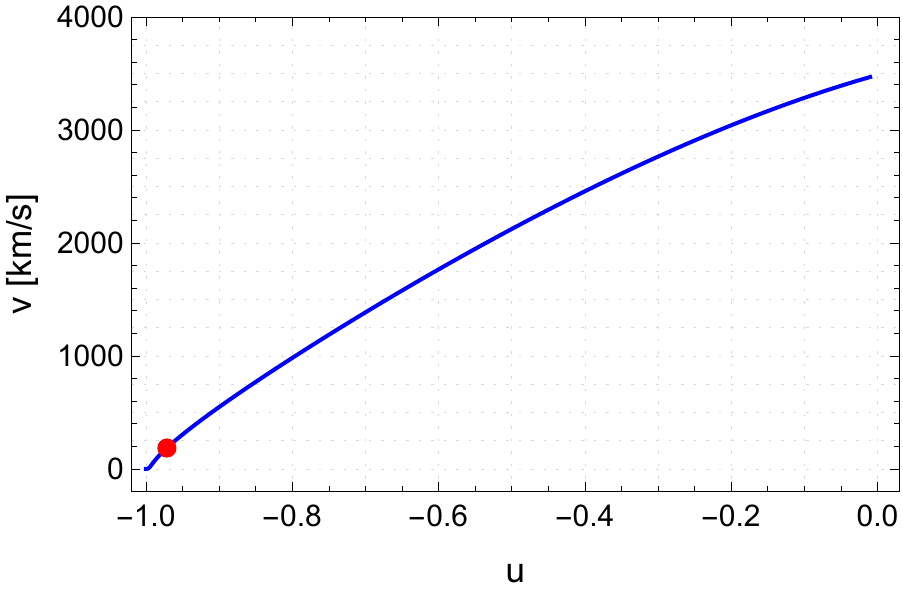}
\caption{Velocity profile for a typical O5 V star without rotation. The velocity profile is plotted as a function of  $\log(r/R_{\ast}-1)$  (left panel) and as a function of $u$ (right panel). The location of the singular point is shown with a red dot, while the sonic point is in black.
\label{velpro}}
\end{figure}

Knowing the topology of the m-CAK model, specifically the locus of singular points, we now solve the EoM for the velocity profile, $v(r)$, or equivalently $w=w(u)$, and the eigenvalue $C'$, which is proportional to the mass loss rate, $\dot{M}$. Then, the wind parameters obtained for this model are: a terminal velocity ($v_{\infty})$ of $3467\, {\rm km/s}$ and a mass-loss rate ($\dot{M}$) of $2.456 \times 10^{-6}\, M_{\odot}/{\rm yr}$. Figure \ref{velpro} shows the velocity profile of this model as a function of $\log(r/R_{*}-1)$ (left panel) and as a function of $u$ (right panel). The location of the singular point ($r_s$) is shown with a red dot, and it is located near the stellar surface ($r_{s}= 1.029 \, R_{*}$ or $u_s=-0.9719$). At this point, the wind velocity is $181.4 \, \rm{km/s}$, a highly supersonic speed ($a= 24.17 \, \rm{km/s})$. 

\begin{figure}[H]
\center
\includegraphics[width=6.8cm]{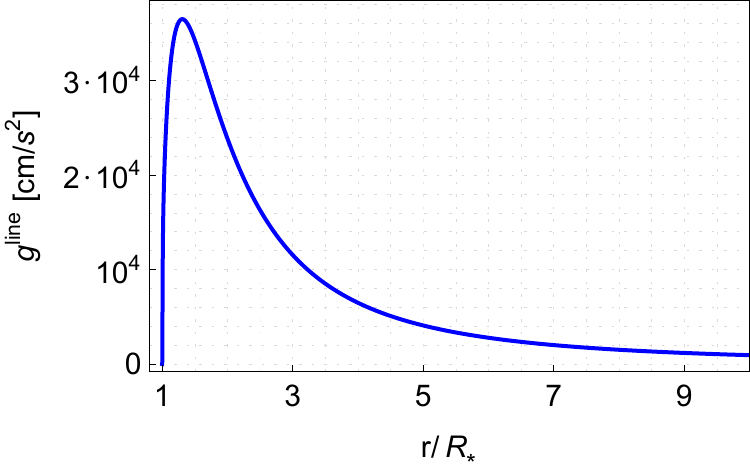}
\caption{The radiative acceleration, $g^{\rm{line}}$, for a typical O5 V star without rotation as function of  $r/R_{*}$, for $r < 10\, R_{\ast}$. 
\label{gline}}
\end{figure}

This steep velocity gradient is due to the rapid increase of the line force just above the stellar surface, as shown in Fig. \ref{gline}, where the sound speed is reached at $r = 1.006\, R_{*}$ and the  maximum of $g^{\rm{line}}$ is reached at $r = 1.3\, R_{*}$.

As previously mentioned, the wind parameters ($v_{\infty}$ and $\dot{M}$) must be calculated within the framework of the radiative transport problem. However, to understand the complex non-linear dependence on the wind parameters from the line-force parameters, in the following figures, we show how the wind parameters depend in terms of each one of the line-force parameters.

\begin{figure}[H]
\center
\includegraphics[width=6.7cm]{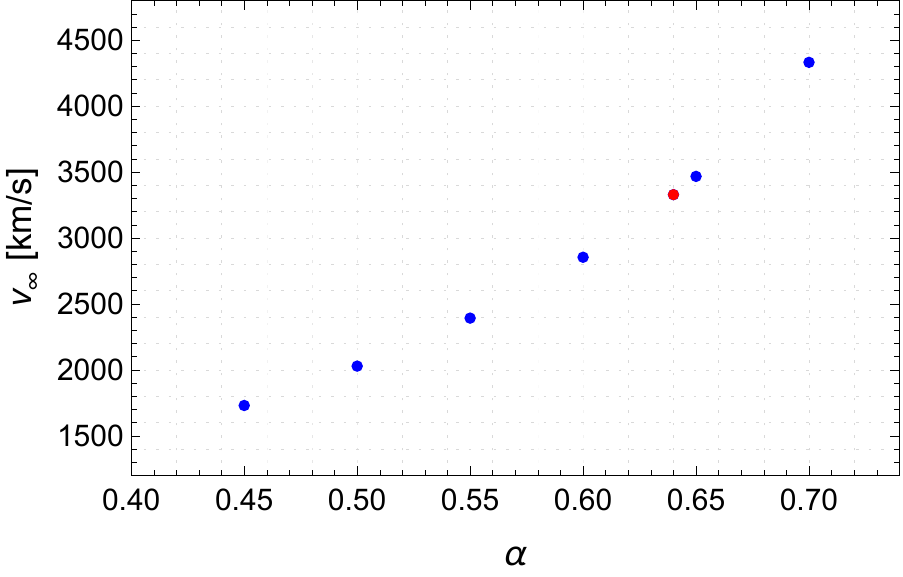}
\includegraphics[width=6.4cm]{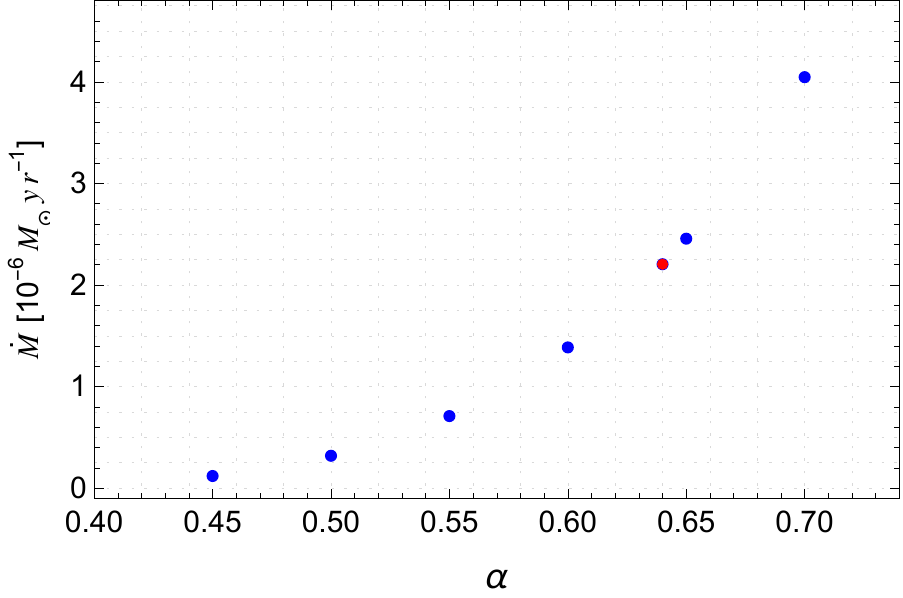}
\caption{Dependence of the wind parameters as a function of the line-force parameter $\alpha$. Terminal velocity (left panel) and mass-loss rate (right panel). The values obtained for our typical O5 V star without rotation are shown in red.
\label{alpha}}
\end{figure}

Figure \ref{alpha} shows the dependence of the wind parameters, $v_{\infty}$ (left panel) and $\dot{M}$ (right panel) as a function of the line force parameter $\alpha$, using the same stellar parameters and keeping the line force parameters $k$ and $\delta$ fixed. There is an increase in the values of both  wind parameters as $\alpha$ increases.

\begin{figure}[H]
\center
\includegraphics[width=6.7cm]{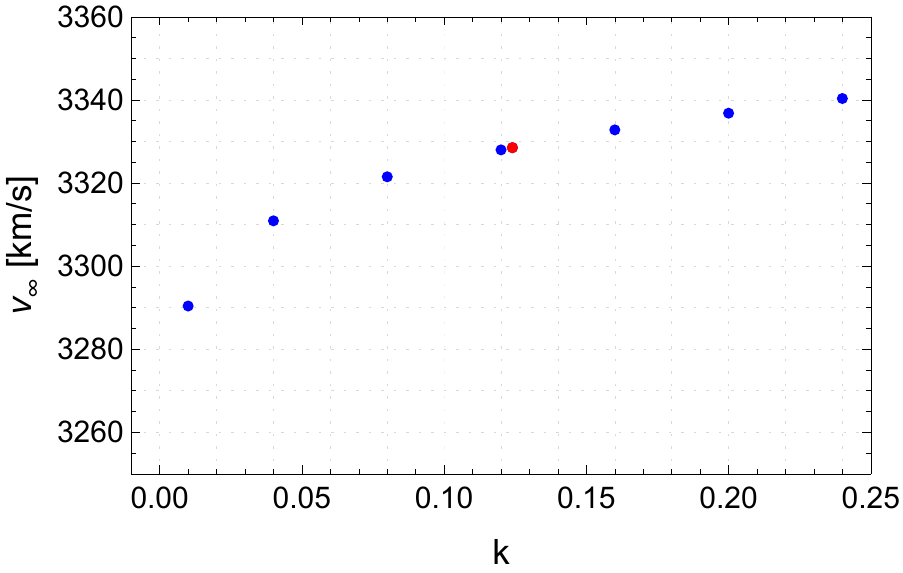}
\includegraphics[width=6.4cm]{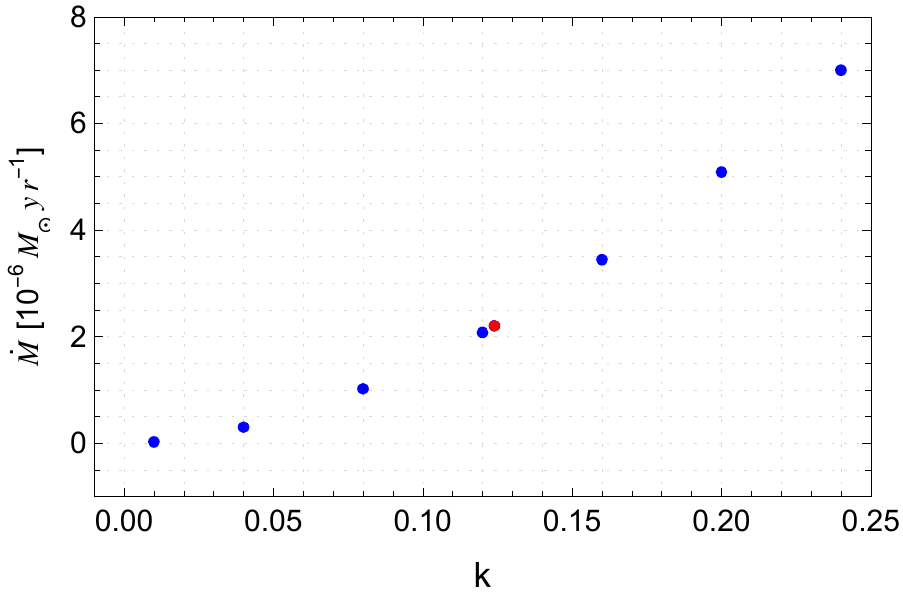}
\caption{Dependence of the wind parameters as a function of the line-force parameter $k$. Terminal velocity (left panel) and mass-loss rate (right panel). The values obtained for our typical O5 V star without rotation are shown in red.
\label{kion}}
\end{figure}

\begin{figure}[H]
\center
\includegraphics[width=6.7cm]{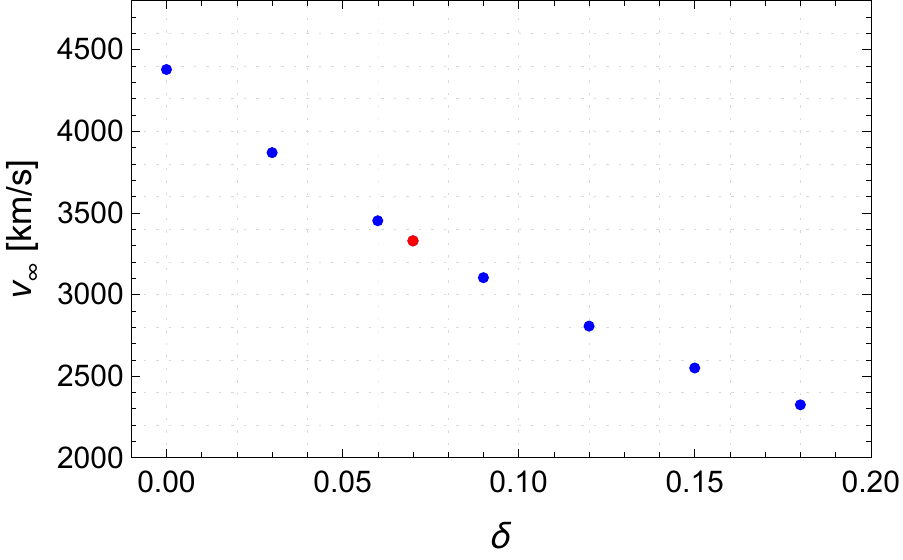}
\includegraphics[width=6.5cm]{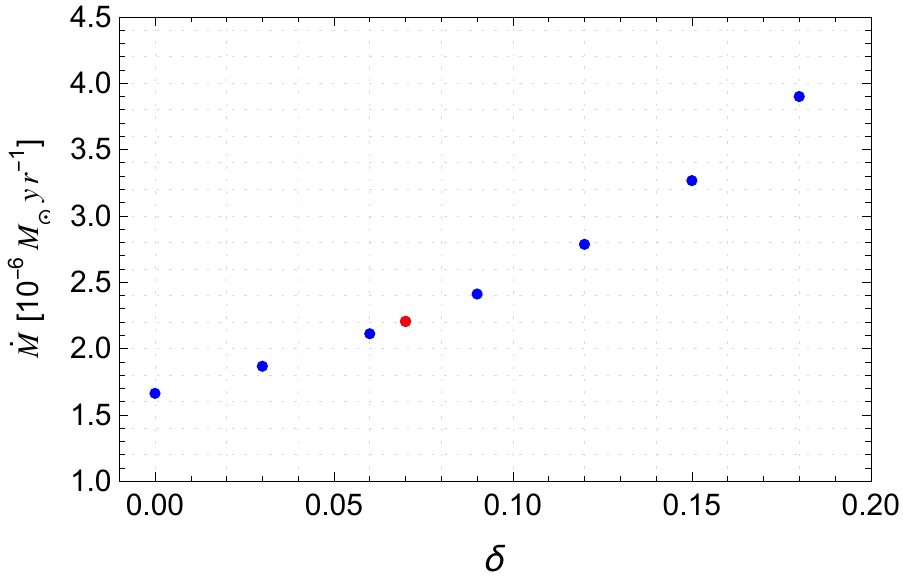}
\includegraphics[width=6.7cm]{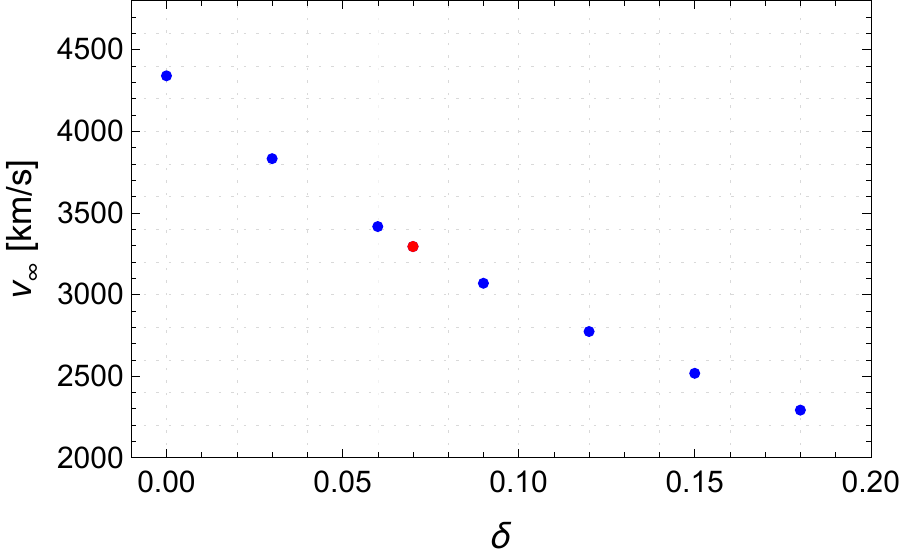}
\includegraphics[width=6.9cm]{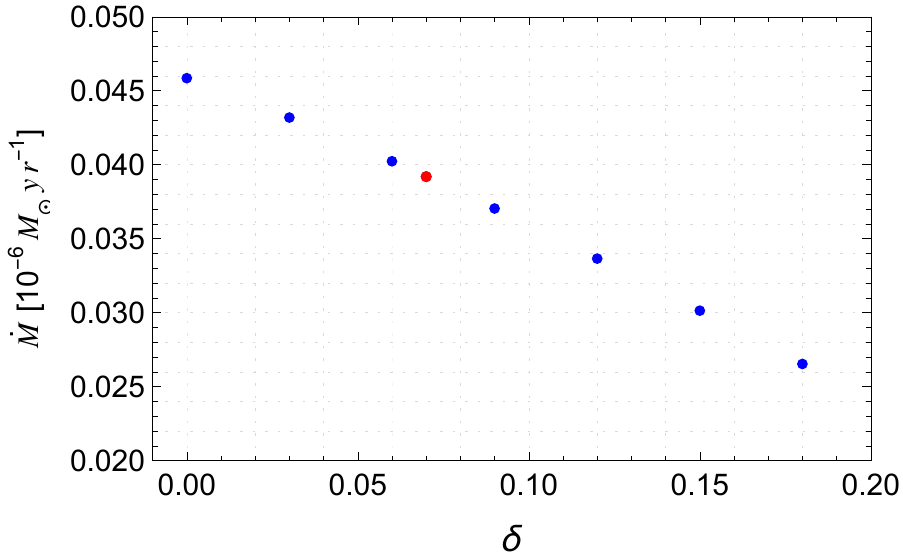}
\caption{Dependence of the wind parameters as a function of the line-force parameter $\delta$. 
Upper panels are for $k=0.124$, and lower panels are for $k=0.0124$. The values obtained for our typical O5 V star without rotation are shown in red.
\label{delta}}
\end{figure}

The dependence of the wind parameters as a function of the line force parameter $k$ is shown in Fig. \ref{kion}. In this case, wind parameters also increase as $k$ increases. It is clearly seen in Fig. \ref{kion} that the terminal velocity depends only slightly on the value of $k$ rather than the mass-loss rate, which has a significant impact on the value of $k$ \citep{venero2016}.
Finally, the dependence of the wind parameters as a function of the line force parameter $\delta$ is shown in Fig. \ref{delta}. We observe that the terminal velocity has a decreasing behaviour  when  the parameter $\delta$ increase, while the mass-loss rate can have a decreasing or increasing behaviour. This behaviour depends on the parameter $k$, for low values of $k$ the mass-loss rate decreases while $\delta$ increases, but for larger values, the behaviour is reversed. 

In the next subsections, we will discuss each of the slow solutions. The combined effect of the line-force parameters and the three physical wind solutions is discussed in detail  in \citet{venero2016}.

Overall, the fast solution from the m-CAK theory has been very successful in explaining the terminal velocities and mass-loss rates of massive stars \cite[see ][]{lamers1999,kudritzki2000,puls2008,vink2022}.

\subsection{$\Omega$-slow solution \label{Oslowsol}}
The original  \citetalias{castor1975} model considered the star point approximation, i.e., all the photons are radially directed over the wind plasma. In that work, \citetalias{castor1975} only discussed the effect of the finite disk of the star seen by an observer in the wind. \citetalias{friend1986} and \citetalias{ppk1986} implemented the finite disk correction factor and solved the EoM. In both works, they also studied the influence of rotation in the equatorial plane of a  rotating star, but they could not obtain solutions for rapidly rotating stars. The reason was found by \citet{cure2004}, for $\Omega \gtrsim 0.75$, where $\Omega=v_{\rm{rot}}/v_{\rm{crit}}$. From this value of $\Omega$, the fast solution ceases to exist, and another type of  solution is found. This solution called the \textit{$\Omega$-slow} solution, is characterised by a slower and denser wind in comparison with the fast solution.

It is well known that Be stars are the fastest rotators among stars \citep{rivinius2013}. Thus, in this section, we will study the topology and the wind solutions for a typical B2.5 V star with the following stellar and line force parameters: $T_{\rm eff}=20000K$, $\log g=4.11$, $R/R_{\odot}=4.0$, $k=0.61$, $\alpha=0.5$, and $\delta=0.0$. The lower (surface) boundary condition is fixed at $\rho_{*}=8.7 \times 10^{-13}\, \mathrm{ g/cm^3}$. In addition, the distortion of the shape of the star caused by its high rotational speed and gravity-darkening effects are not considered.

\begin{figure}[H]
\center
\includegraphics[width=6.8cm]{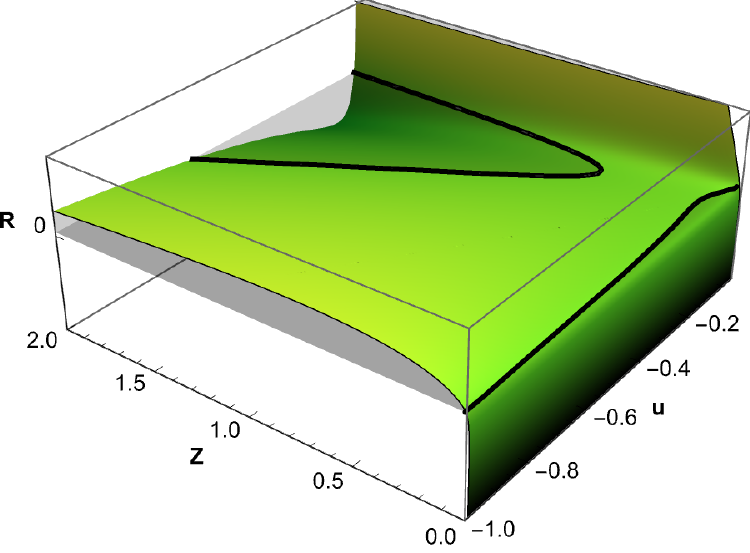}
\includegraphics[width=6.8cm]{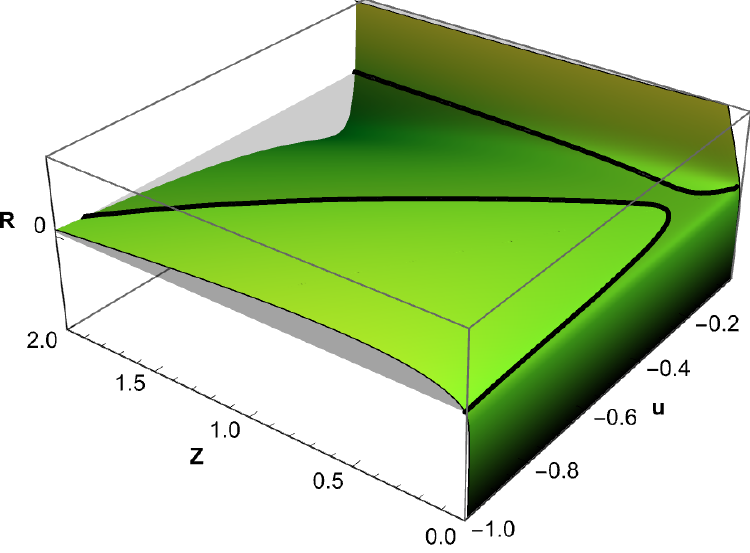}
\includegraphics[width=6.8cm]{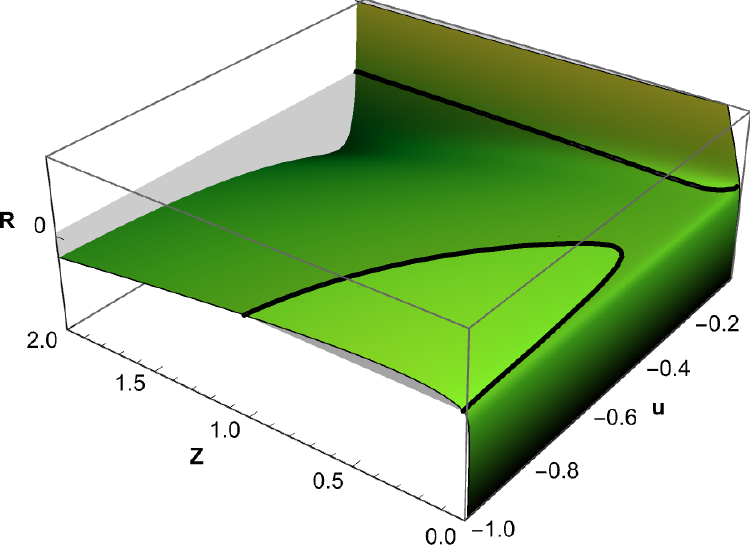}
\includegraphics[width=6.8cm]{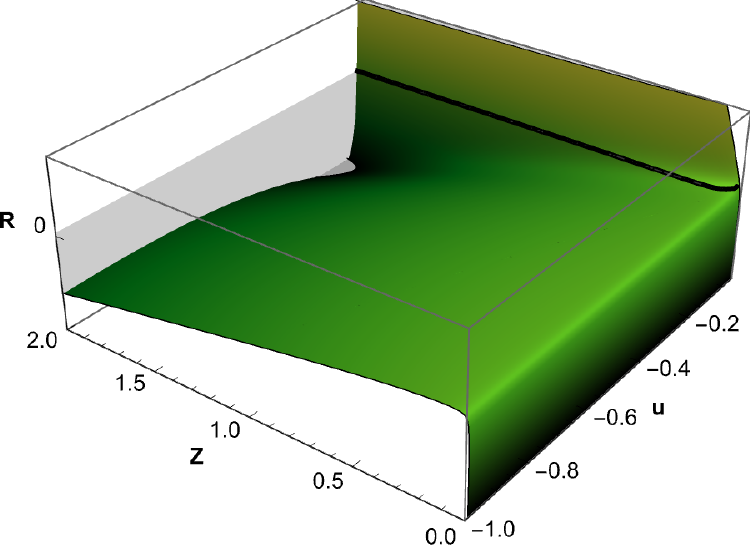}
\caption{The function $R(u,Z)$ (topology) of the m-CAK theory as function of $\Omega$: $\Omega=0.3$ (upper left panel), $\Omega=0.5$ (upper right panel), $\Omega=0.7$ (lower left panel), and $\Omega=0.9$ (lower right panel). The plane $R(u,Z)=0$ is shown in light grey, and its intersection with the surface $R(u,Z)$ (locus of singular points) is plotted with black lines. 
\label{fig5}}
\end{figure}

Fig. \ref{fig5} shows the surfaces $R(u,Z)$, for different values of $\Omega$, together with the plane $R(u,Z)=0$. The intersection of the surfaces $R(u,Z)$ and $R(u,Z)=0$ (black lines) correspond to the locus of singular points. We clearly  observe two different loci of singular points. The fast solution locus can be observed for $\Omega=0.3$ (upper left panel), $\Omega=0.5$ (upper right panel) and $\Omega=0.7$ (lower left panel). For larger rotational rates  ($\Omega \gtrsim 0.75$), the fast solution locus lies completely under the plane $R(u,Z)=0$, as shown in the lower right panel for $\Omega=0.9$. Thus, the fast solution \textit{does not exist} for large values of $\Omega$.

\begin{figure}[H]
\center
\includegraphics[width=6.8cm]{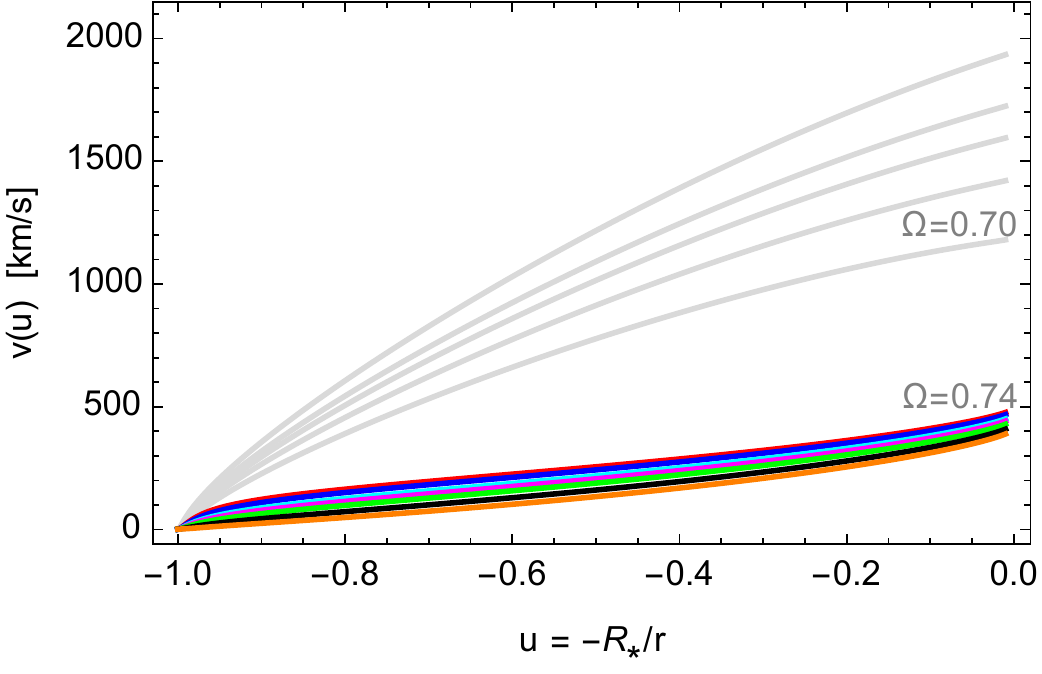}
\includegraphics[width=6.8cm]{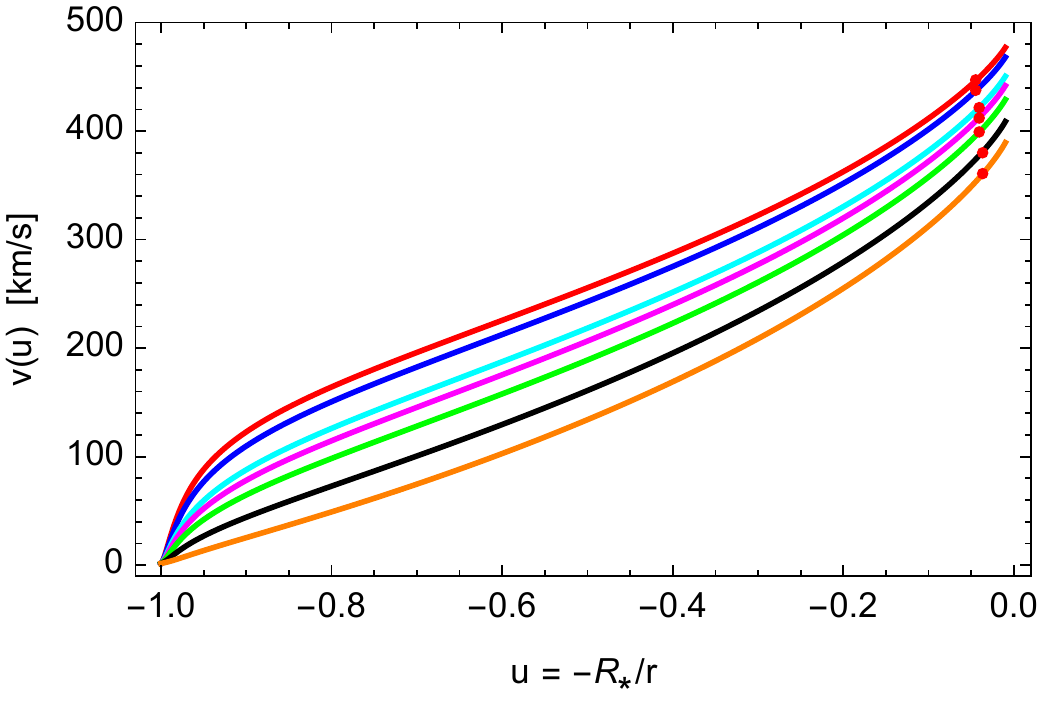}
\caption{Velocity profiles, $v(u)$, as function of the rotational rate $\Omega$. Left panel: fast solutions ($\Omega \lesssim 0.74$) are plotted in grey lines, while the $\Omega$-slow solutions are in coloured lines: $\Omega= 0.74$ (red line), $\Omega= 0.76$ (blue line), $\Omega= 0.80$ (cyan line), $\Omega= 0.82$ (magenta line), $\Omega= 0.85$ (green line), $\Omega= 0.90$ (black line), and $\Omega= 0.95$ (orange line). Right panel: The same $\Omega$-slow solutions, but zoomed and including the location of the singular points (red dots).
\label{fig6}}
\end{figure}  

Figure \ref{fig6} shows the velocity profiles, $v(u)$ as a function of $u$ for different values of $\Omega$. All these solutions use the same lower boundary condition. This figure shows (left panel) fast solutions in light grey and $\Omega$-slow solutions in coloured lines. The right panel shows only $\Omega$-slow solutions; the location of the singular point is almost independent of $\Omega$. This is a consequence of the shape of the locus curve of singular points (see Fig. \ref{fig5}). This locus is located almost at a constant value of $u$.

\begin{figure}[H]
\center
\includegraphics[width=6.8cm]{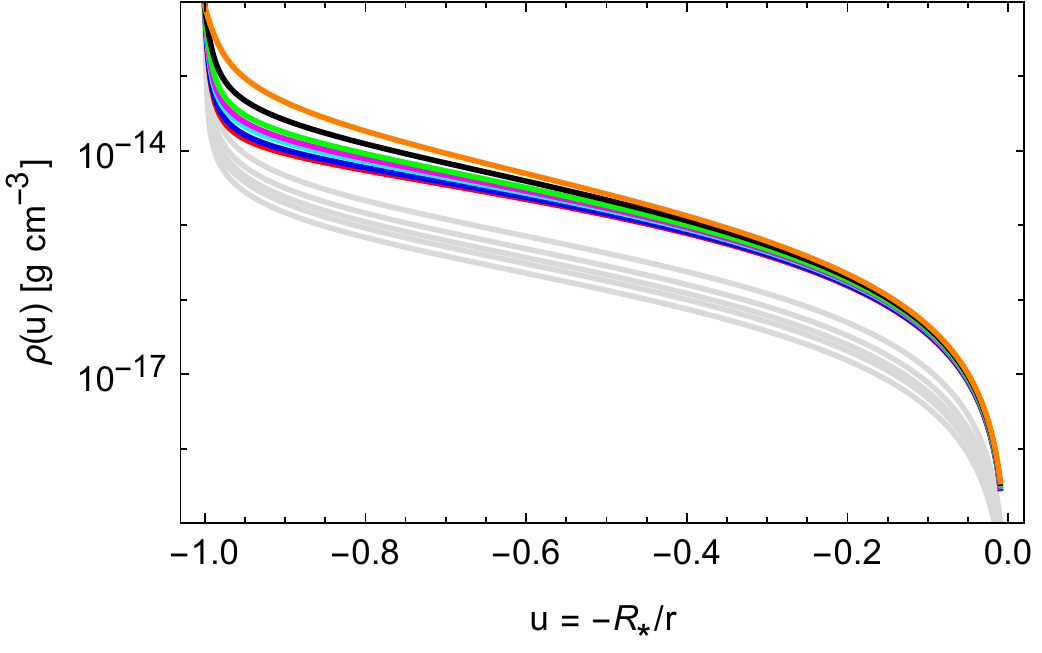}
\caption{Wind density profiles, $\rho(u)$ (in gr/cm$^3$) versus $u$, 
for fast and $\Omega$-slow solutions. The colour scheme is the same as the one used in Fig. \ref{fig6} \label{fig7}}
\end{figure}

The $\Omega$-slow solutions are only valid in the equatorial plane in this 1D m-CAK model. Notice that this model does not take into account the oblateness and gravity-darkening effects. See \citet{araya2017} for the implementation in the 1D model (equatorial plane)  and \citet{cranmer1995} for the implementation in the 2D model. \\ 
In the equatorial plane, the higher  $\Omega$, the greater the centrifugal force and, consequently, the lesser the effective gravity. Therefore, the higher $\Omega$ is, the higher the rate of mass loss and, through the continuity equation, the higher the wind density, as shown in Fig.\, \ref{fig7}.

\subsection{$\delta$-slow solution \label{Dslowsol}}

The \textit{$\delta$-slow} solution was found numerically by \citet{cure2011}. 
This solution, based on the m-CAK  theory, describes the wind velocity profile when the ionization-related line-force parameter $\delta$ takes larger values, $\delta \gtrsim  0.28$. These values are larger than the ones provided by the standard m-CAK solution \citep[see,][and references therein]{lamers1999}. Nevertheless, \citet{puls2000} calculated the value of $\delta$ for a pure Hydrogen atmosphere finding a value of $\delta\sim 1/3$. 
These high values of $\delta$ are also found in atmospheres and winds with extremely low metallicities 
\citep[see,][]{kudritzki2002}.
 
The $\delta$-slow solution, as well as the $\Omega$-slow solution, is characterized by low velocities. This solution could explain the  velocities obtained for late-B and A-type supergiant stars and seems to fit  well the observed anomalous correlation between the terminal and escape velocities found in A supergiant stars \citep{cure2011}. 
Furthermore, in \citet[][see their table 2]{venero2016}, a gap of solutions, between the fast and the $\delta$-slow solutions, for different values of the rotational speed, was found in the plane $\delta$-$\Omega$. 

To present the topological analysis of this type of solution, we adopt the model T19 from \citet{venero2016}. The stellar and line-force parameters are $T_{\rm eff}=19000K$, $\log g=2.5$, $R/R_{\odot}=40$, $k=0.32$, and $\alpha=0.5$.  We use $\tau_{*}=2/3$ as a boundary condition at the stellar surface.

\begin{figure}[H]
\center
\includegraphics[width=6.8cm]{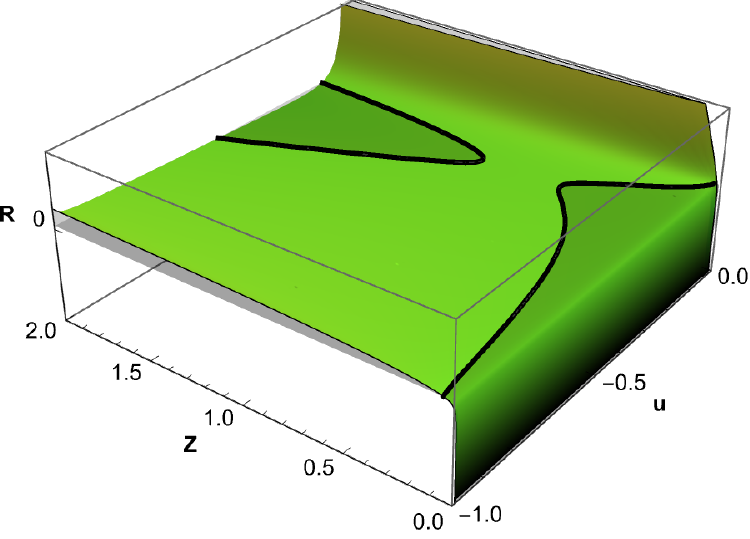}
\includegraphics[width=6.8cm]{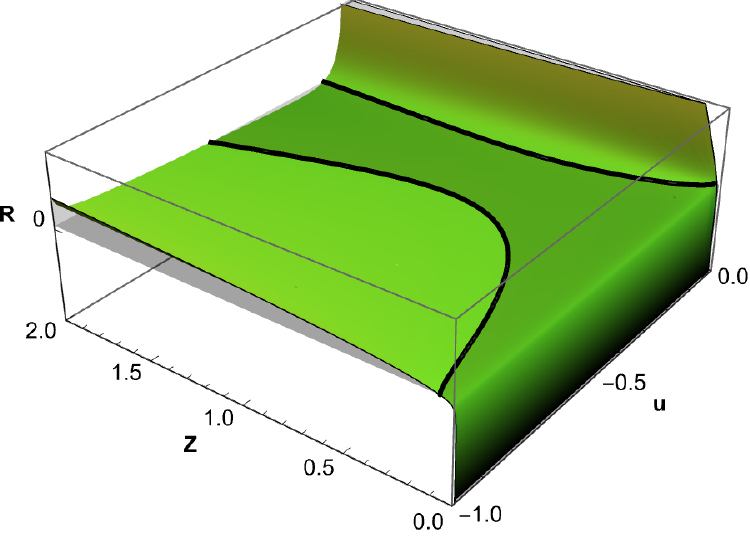}\\
\includegraphics[width=6.8cm]{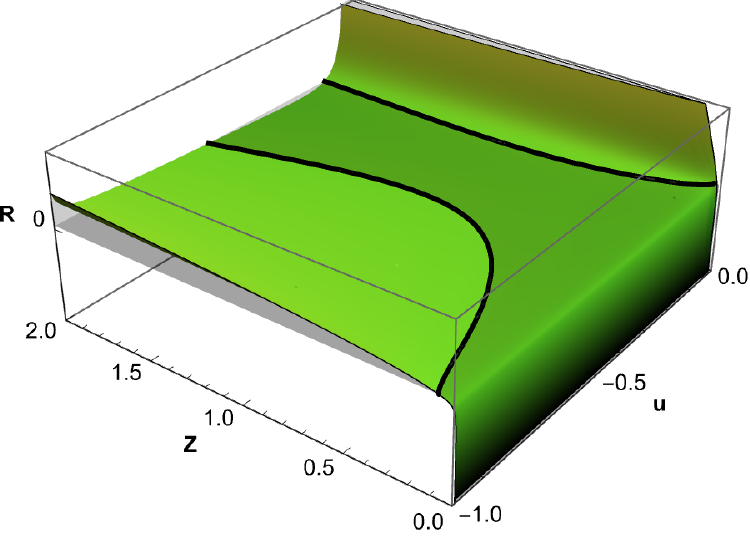}
\includegraphics[width=6.8cm]{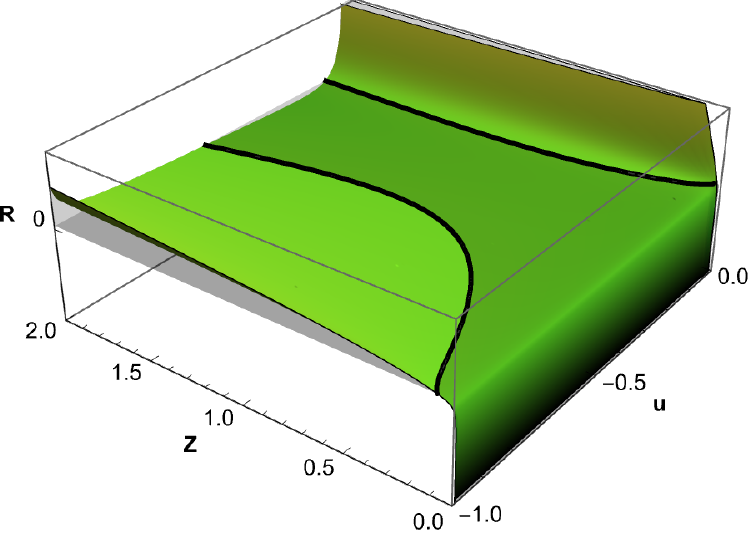}
\caption{The topological function $R(u,Z)$ of the m-CAK theory as function of $\delta$. Upper left panel: $\delta=0.1$. Upper right panel: $\delta=0.12$. Lower left panel: $\delta=0.24$. Lower right panel: $\delta=0.3$. The plane $R(u,Z)=0$ is shown in light grey, and its intersection with the surface $R(u,Z)$ (locus of singular points) is plotted with black lines. 
\label{fig8}}
\end{figure}

In Fig. \ref{fig8}, the $R(u,Z)$ function and the plane $R(u,Z)=0$ are shown for different values of $\delta$. The upper left panel shows the surface $R(u,Z)$ for $\delta=0.1$. We clearly see that for this case, the locus of singular points for fast solutions is different from the case of the fast solution shown in Fig. \ref{fig1}. Here this locus is located when $Z \lesssim 0.5$, $\forall\, u$. The upper right panel shows $R(u,Z)$ for $\delta=0.12$, where the locus of singular points for fast solutions returns to the behaviour shown in Fig. \ref{fig1}. The fast solution is present until $\delta=0.24$; see the lower left panel. We cannot find fast solutions for slightly larger values of $\delta$ until $\delta \sim 0.30$ (lower right panel). For this value of $\delta$, the locus of singular points for fast solutions shifts to slightly larger values of $Z$ for $u \lesssim -1$, and the numerical wind solutions no longer have a singular point in this locus, switching to the other locus of singular points ($\delta$-slow solutions) located at $u \lesssim -0.1$ (or $r \gtrsim 10 R_{*}$).

\begin{figure}[H]
\center
\includegraphics[width=6.8cm]{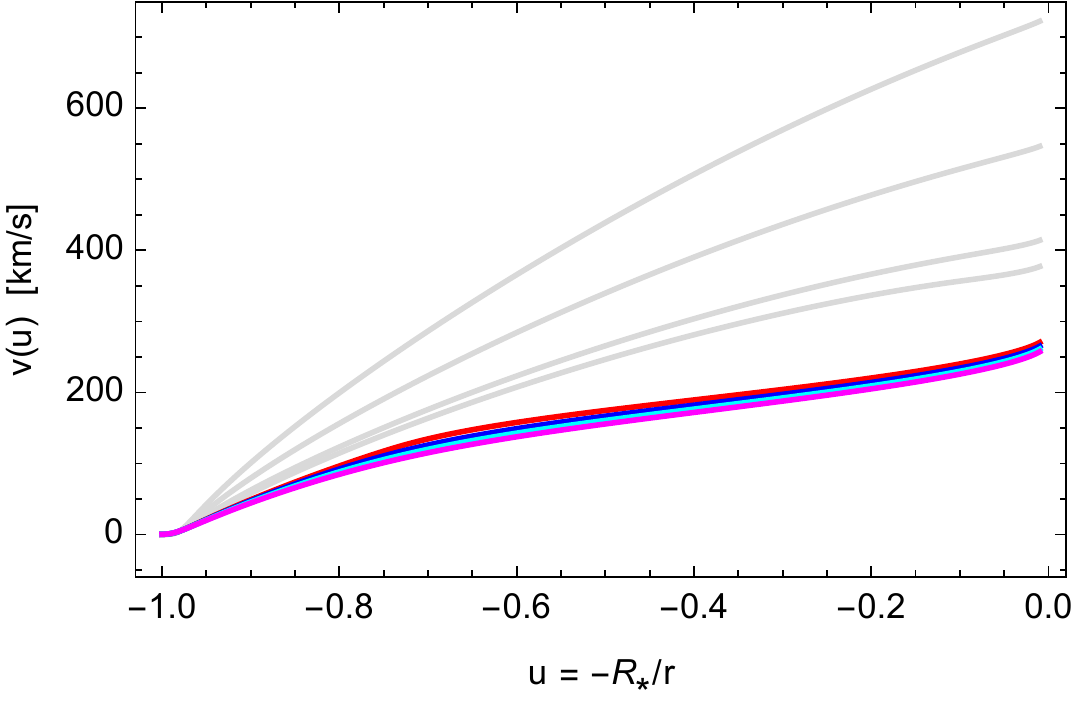}
\includegraphics[width=6.8cm]{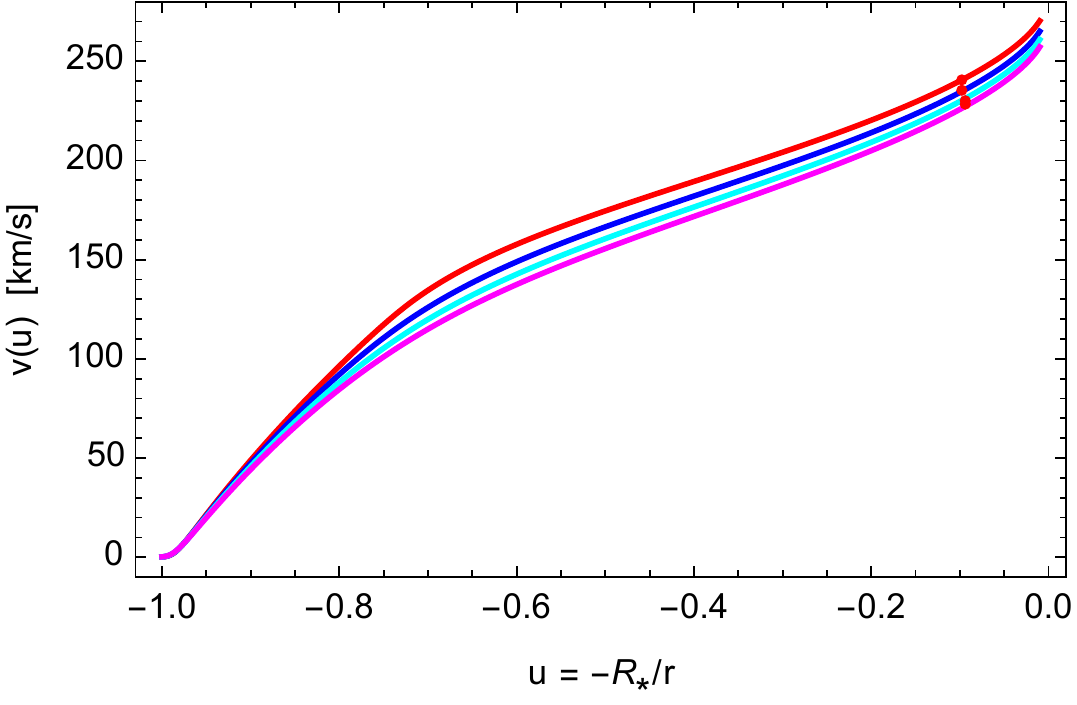}
\caption{Velocity profiles, $v(u)$, for different values of the line-force parameter $\delta$. Left panel: fast solutions are plotted in grey lines for $\delta=0.0,0.1,0.2,0.24$, while $\delta$-slow solutions are in coloured lines. Red line corresponds to $\delta= 0.3$, blue line to $\delta= 0.31$, cyan line to $\delta= 0.32$, and magenta line to $\delta= 0.33$. The right panel shows only  $\delta$-slow solutions, and the location of the singular points for each solution is shown with a red dot.
\label{fig9}}
\end{figure}  

\subsection{The $\beta$-law approximation \label{betalaw}}
In the work of \citetalias{ppk1986}, after obtaining the numerical solution of the EoM, they assumed a power law approximation to describe the velocity profile only as a function of the radial coordinate $r$. This approximation is known as the \textit{$\beta$-law} approximation and has the following expression:
\begin{eqnarray}
v(r) &=& v_{\infty} \, (1-R_{\ast}/r)^\beta \, ,\\
 &=& v_{\infty} \, (1+u)^\beta \,,
\end{eqnarray}
where $v_{\infty}$ is the terminal velocity and the value of $\beta$ determine the shape of the velocity profile. In the context of stellar wind diagnostics, these parameters are considered fit parameters that must be determined through spectral line fitting. Usually, the range  used for the $\beta$ parameter is  $0.7 \lesssim \beta \lesssim 4 $ \citep{kudritzki2000}. 

\begin{figure}[H]
\center
\includegraphics[width=6.8cm]{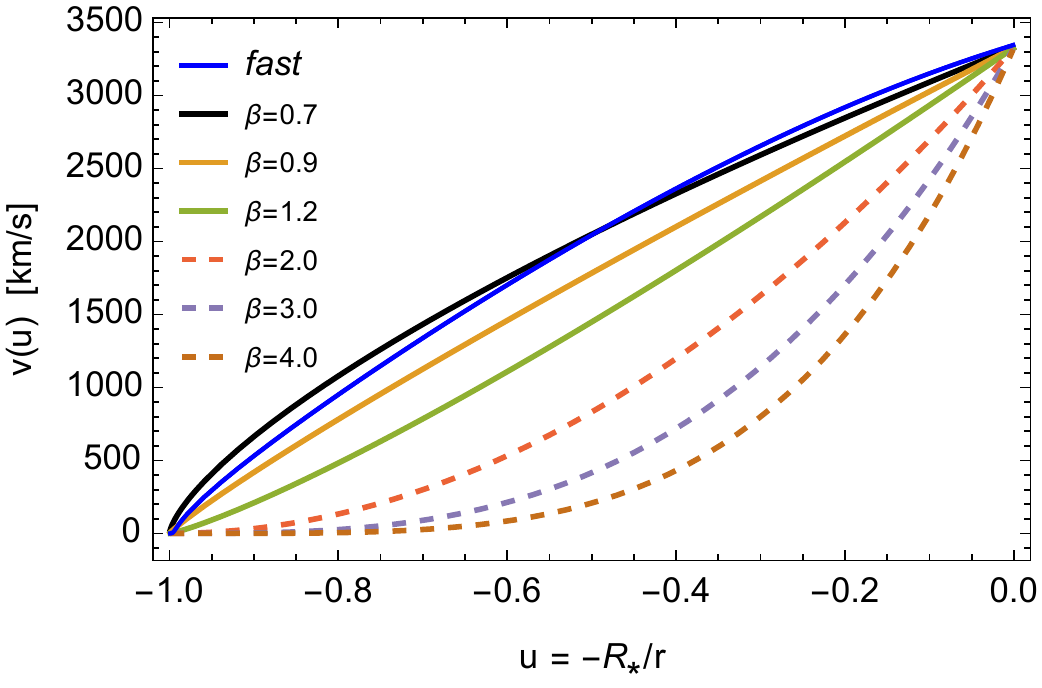}
\caption{Fast solution velocity profile (solid blue), $v(u)$ vs. $u$. Six different $\beta$-law velocity profiles are also plotted. It is clearly seen that the $\beta$-law approximation is a good one for  $0.7 \lesssim \beta \lesssim 1.2$. See text for details.
\label{fig10}}
\end{figure}

Figure \ref{fig10} shows the velocity profile of the fast solution for the stellar and line-force parameters given at the beginning of section \ref{fastsol}, together with  six different values of $\beta$ for a $\beta$-law velocity profile. From this figure, we can conclude that the fast solution cannot be described properly by a $\beta$-law with $\beta > 1.2$.

\begin{figure}[H]
\center
\includegraphics[width=6.8cm]{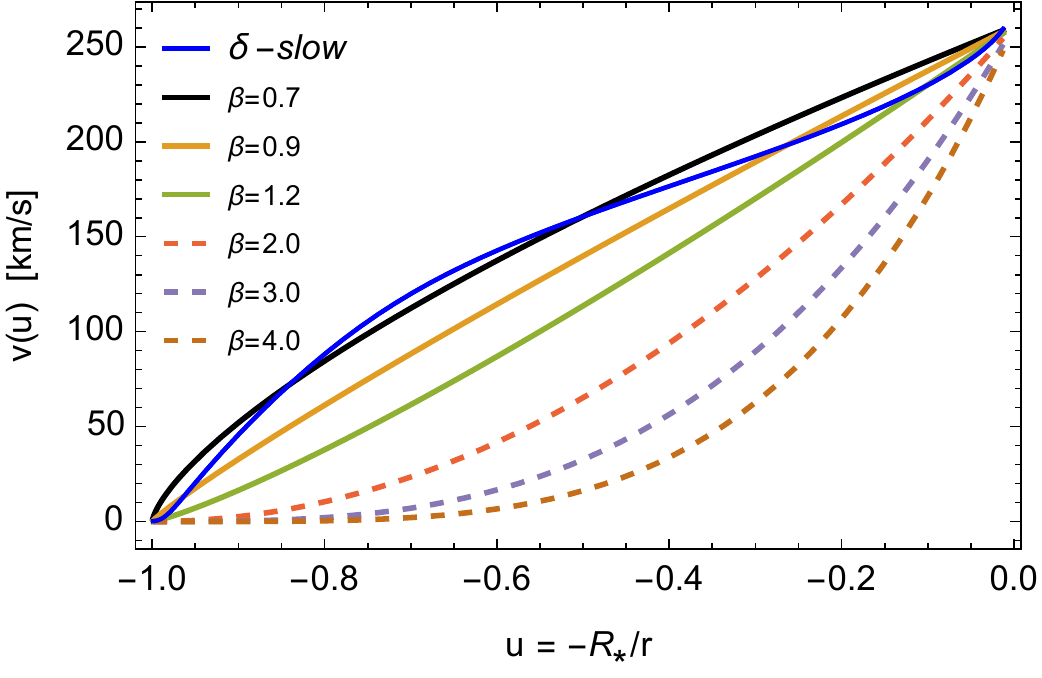}
\caption{$\delta$-slow solution velocity profile (solid blue), $v(u)$ vs. $u$. Six different $\beta$-law velocity profiles are also plotted. For values around to $\beta=0.7$ the profiles can be considered similar, but it can be clearly concluded that for $\beta>0.7$, the $\beta$-law profile cannot fit the m-CAK hydrodynamical $\delta$-slow solution.
\label{fig11}}
\end{figure}

\begin{figure}[H]
\center
\includegraphics[width=6.8cm]{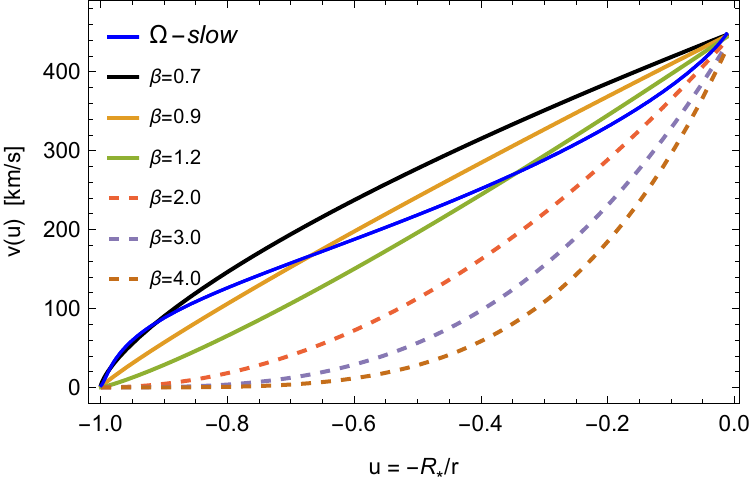}
\caption{$\Omega$-slow solution velocity profile (solid blue), $v(u)$ vs. $u$. Six different $\beta$-law velocity profiles are also plotted. It can be clearly concluded that no $\beta$-law profile can fit the m-CAK hydrodynamical $\Omega$-slow solution.
\label{fig12}}
\end{figure}

On the other hand, Figure \ref{fig11} shows the velocity profile for the $\delta$-slow solution for the stellar and line-force parameters given at the beginning of section \ref{Dslowsol}, and $\delta=0.32$. Also, the same $\beta$-law profiles of Fig. \ref{fig10} are used, with the proper values of $v_{\infty}$ for this solution. The same is plotted in Fig. \ref{fig12} for the $\Omega$-slow solution with a $\Omega=0.8$. 
In the case of the $\delta$-slow solution, the $\beta$-law profile cannot fit the m-CAK hydrodynamical solution for any $\beta>0.7$, while for values around to $\beta=0.7$ the profiles can be considered similar. Finally, from Fig. \ref{fig12}, we can definitely conclude that $\Omega$-slow solutions cannot be described properly by any $\beta$-law profile.

\section{Analytical wind solutions \label{analyticalsols}}

Using an analytical expression to represent the radiation force and solve the equation of motion analytically offers numerous advantages over the numerical integration of the EoM. These formulae can be used  in all cases where good first estimates are needed; for example, it gives the advantage of solving the radiative transfer  problem for moving media in an easier way.

 \citetalias{kppa1989} were one of the first to develop analytical solutions for  radiation-driven winds considering the finite disk correction factor in the line force. Based on these solutions, they provided approximated analytical expressions for the terminal velocity and the mass loss rate in terms of the stellar parameters ($L$, $M_{\ast}$ and $R_{\ast}$), the line force parameters ($k$, $\alpha$ and $\delta$) and the free parameter $\beta$ from the $\beta$-law (they adopted $\beta=1.0$ for their results). 

Other authors have tried to simplify the complicated numerical treatment from the theory. \citet{villata1992} (hereafter V92), with the purpose of simplifying the integration of the EoM, derived an approximated expression for the line acceleration term, which depends only on the radial coordinate. \citet{muller2008} (hereafter MV08) presented an analytical expression for  the  velocity  field  using  a  parameterized  description  for the line acceleration that (as in \citetalias{villata1992}’s) also depends on the radial  coordinate. These line-acceleration expressions do not depend on the velocity or the velocity gradient, as the standard  m-CAK  description  does.
\citet{araya2014} proposed to achieve a complete analytical description of the 1-D hydrodynamical solution for radiation-driven winds in the fast regime  by gathering the advantages of both previous approximations (the use of known parameters and the Lambert$W$ function). In addition, \citet{araya2021} developed an analytical solution for the $\delta$-slow regime. To date, no approximation using the Lambert$W$ function has been performed for the $\Omega$-slow regime, we expect to do this in the  future.

In the next sections, we describe the results and methodology used to solve analytically the equation of motion for fast and $\delta$-slow regimes.

\subsection{Solution of the dimensionless equation of motion}

In this section, we recapitulated the methodology described by \citetalias{muller2008} to obtain the dimensionless equation of motion.

In a dimensionless form, the momentum equation can be expressed as follows, 

\begin{equation}
\hat{v} \, \frac{d \hat{v}}{d \hat{r}}= -\frac{\hat{v}_{\rm{crit}}^{2}}{\hat{r}^{2}} + \hat{g}^{\rm{line}} - \frac{1}{\rho}\frac{d \rho}{d \hat{r}},
\end{equation}

\noindent where $\hat{r}$ is a dimensionless radial coordinate $\hat{r}=r/R_{*}$, and the dimensionless velocities 
(in units of the isothermal sound speed $a$) are:

\begin{equation}
\hat{v}=\frac{v}{a}\:\:\:\:\: \mathrm{and} \:\:\:\:\: \hat{v}_{\rm{crit}}=\frac{v_{\rm{esc}}}{a\sqrt{2}}, ,
\end{equation}
here, $v_{\rm{crit}}$ is the rotational break-up velocity in the case of a rotating star. It is usually determined by dividing the effective escape velocity, $v_{\rm{esc}}$,  by a factor of $\sqrt{2}$. Similarly, a dimensionless line acceleration can be written as follows:

\begin{equation}
\label{norma}
\hat{g}^{\rm{line}}=\frac{R_{*}}{a^{2}}\, g^{\rm{line}}.
\end{equation}

\noindent Using the continuity equation and the equation of state of an ideal gas, the dimensionless equation of motion reads as follows:

\begin{equation}
\label{motion}
\left( \hat{v} - \frac{1}{\hat{v}} \right) \frac{d\hat{v}}{d \hat{r}}= -\frac{\hat{v}_{\rm{crit}}^{2}}{\hat{r}^{2}} + \frac{2}{\hat{r}} + \hat{g}^{\rm{line}} (\hat{r}).
\end{equation}

Lastly, a 1-D velocity profile is derived analytically in terms of Lambert$W$ function \citep{corless1993,corless1996,cranmer2004}. See \citetalias{muller2008} for a detailed description of the  methodology used to arrive at this solution. This analytical solution is expressed as follows: 
\begin{equation}
\label{V-profile}
\hat{v}(\hat{r})= \sqrt{-W_{j}(x(\hat{r}))},
\end{equation}
where
\begin{equation}
\label{eq-X}
x(\hat{r})= -\left(  \frac{\hat{r}_{\rm c}}{\hat{r}} \right) ^{4}  \, \exp \left[ -  2 \, \hat{v}^{2}_{\rm{crit}} \left( \frac{1}{\hat{r}} - \frac{1}{\hat{r}_{\rm c}}  \right)
-2 \int_{\hat{r}_{\rm c}}^{\hat{r}} \hat{g}^{\rm{line}}(\hat{r}) d\hat{r} - 1 \right]\, .
\end{equation}

\noindent In the last equation appears the parameter $\hat{r}_{\rm c}$, which represents the position of the sonic (or critical) point. 

Also, the Lambert$W$ function ($W_j$) has only two real branches, indicated by the sub-index {\it j}, where $j=0,-1$. These two branches coincide at the sonic point, $\hat{r_{\rm c}}$, i.e., 
\begin{equation}
j = \left\{  \begin{array}{r r c}
    0 & \mathrm{for} &  1\leq \hat{r} \leq \hat{r}_{\rm c}\\
    -1 & \mathrm{for} & \hat{r}>\hat{r}_{\rm c},
  \end{array} \right. 
\end{equation}

A regularity condition must be imposed, as in the m-CAK case, since the LHS of the equation of motion (Eq. \ref{motion}) vanishes at $\hat{v}=1$ (singularity condition in CAK formalism). This is equivalent to ensuring that the RHS of Eq. (\ref{motion}) vanishes at $\hat{r}=\hat{r}_{\rm c}$. 
Therefore, 
\begin{equation}
-\frac{\hat{v}_{\rm{crit}}^{2}}{\hat{r}_{c}^{2}} + \frac{2}{\hat{r}_{c}} + \hat{g}^{\rm{line}}(\hat{r}_{c}) =0,
\end{equation}
\noindent and $\hat{r}_{\rm c}$ is obtained by solving this last equation. Finally, the velocity profile is derived using the function $x(\hat{r})$, Eq. (\ref{eq-X}), into  Eq. (\ref{V-profile}). 
\subsection{The fast regime}

\citetalias{kppa1989}'s analytical study of radiation-driven stellar winds allowed \citetalias{villata1992} to derive an approximate expression for the line acceleration term. In this case, the line acceleration is only dependent on the radial coordinate, and it reads as follows: 

\begin{equation}
g^{\mathrm{line}}_{\mathrm{V92}}(\hat{r})=\frac{G\,M_{*}\,(1-\Gamma_{e})}{R_{*}^{2}\,\hat{r}^{2}}\,A(\alpha, \beta, \delta) \left( 1- \frac{1}{\hat{r}}\right)^{\alpha (2.2 \, \beta -1)}\, ,
\label{villata-gline}
\end{equation}
with
\begin{eqnarray}
\label{villata-gline2}
A(\alpha, \beta, \delta)=\frac{(1.76\, \beta)^{\alpha}}{1-\alpha}\left[10^{-\delta}(1+\alpha)\right]^{1/(1-\alpha)} 
\left[ 1 + \left( \frac{2}{\alpha} \left\lbrace 1- \left[ 10^{-\delta} (1+\alpha)\right] ^{1/(\alpha -1)} \right\rbrace \right) ^{1/2} \right] ^{\alpha} .
\end{eqnarray}

\noindent According to \citet{kpp1987}, the exponent $\beta$ can be calculated based on the force multiplier parameters and the escape velocity, $v_{\rm{esc}}$:

\begin{equation}
\beta = 0.95 \, \alpha + \frac{0.008}{\delta}+\frac{0.032 \, v_{\rm{esc}}}{500} , 
\end{equation}

\noindent with $v_{\rm{esc}}$ in km/s.

Then, using Eq. (\ref{villata-gline}) in its dimensionless form (Eq. \ref{norma}) and inserting it into the  dimensionless equation of motion (Eq. \ref{motion}), it yields:

\begin{equation}
\label{villata-motion}
\left( \hat{v} - \frac{1}{\hat{v}} \right) \frac{d\hat{v}}{d \hat{r}} = -\frac{\hat{v}_{\rm{crit}}^{2}}{\hat{r}^{2}} + \frac{2}{\hat{r}}  
+ \frac{1}{a^{2}} \frac{GM_{*} (1-\Gamma_{e})}{R_{*}\,\hat{r}^{2}}\, A(\alpha, \beta, \delta) \left( 1- \frac{1}{\hat{r}}\right)^{\gamma_{\rm v}},
\end{equation}
\\

\noindent with $\gamma_{\rm v}\,=\,{\alpha \, (2.2 \, \beta -1)}.$

Based on \citetalias{villata1992}'s approximation of the line acceleration, this differential equation can be viewed as a solar-like differential equation of motion. Hence, the singular point is the sonic point. Additionally, \citetalias{villata1992}'s equation of motion does not have eigenvalues, which means it doesn't depend explicitly on the star's mass loss rate.

Using a standard numerical integration method, \citetalias{villata1992} solved the equation of motion and obtained terminal velocities that were within 3-4 $\%$ of those computed by \citetalias{ppk1986} and \citetalias{kpp1987}.

A parametrized description of line acceleration was presented years later by \citetalias{muller2008}  that is dependent on the radial coordinate (like \citetalias{villata1992}'s). The line acceleration in \citetalias{muller2008} was determined using Monte-Carlo multi-line radiative transfer calculations \citep{koter1997,vink1999} and a $\beta$ law. Following this, the 
line acceleration was fitted using the following formula:
\begin{equation}
\label{MV-gline}
\hat{g}^{\mathrm{line}}_{\mathrm{MV08}}(\hat{r})= \frac{\hat{g}_{0}}{\hat{r}^{1+ \delta_{1}}} \left(  1-\frac{\hat{r_{0}}}{\hat{r}^{\delta_{1}}} \right) ^{\gamma},
\end{equation}
where $\hat{g_{0}}$, $\delta_{1}$, $\hat{r_{0}}$ and $\gamma$ are the acceleration line parameters.

Then, the solution of the equation of motion, based  on  their methodology and line acceleration, is: 
\begin{equation}
\hat{v}(\hat{r})= \sqrt{-W_{j}(x(\hat{r}))}\, ,
\end{equation}
with 
\begin{align}
x(\hat{r})=& -\left(\frac{\hat{r}_{\rm c}}{\hat{r}} \right)^{4}  \exp \left[ -  2 \, \hat{v}^{2}_{\rm{crit}} \left( \frac{1}{\hat{r}} - \frac{1}{\hat{r}_{\rm c}}  \right) \right. \nonumber \\
&-\frac{2}{\hat{r}_{0}} \frac{\hat{g}_{0}}{\delta_{1} \, (1+ \gamma)}  \left( \left. \left( 1- \frac{\hat{r}_{0}}{\hat{r}^{\delta_{1}}} \right) ^{1+\gamma} - \left( 1- \frac{\hat{r}_{0}}{\hat{r}_{\rm c}^{\delta_{1}}} \right) ^{1+\gamma} \right)  - 1 \right].
\end{align}

As a result of the approximations described above, the velocity profile can be represented analytically, greatly simplifying the solution of the equation of motion.

Also, it is relevant to note that each of the mentioned approximations has its own advantages and disadvantages. Even though Villata's approximation of the radiation force is general and can directly be applied to describe any massive star's wind, the momentum equation still needs to be solved numerically. With \citetalias{muller2008}'s approximation, the equation of motion can be analytically solved, based on $\hat{g}_{0}$, $\delta_{1}$, $\hat{r}_{0}$, and $\gamma$ parameters of the star. Nevertheless, it is still necessary to perform Monte Carlo multi-line radiative transfer calculations in order to determine these parameters.

This methodology to solve the equation of motion  was revisited by  \citet{araya2014}, and in order to derive a fully analytical expression combining \citetalias{villata1992}'s expression of the equation of motion, with the methodology developed by \citetalias{muller2008}.\\ 
This analytical solution is,
\begin{equation}
\label{sol-V-MV}
\hat{v}(\hat{r})= \sqrt{-W_{j}(x(\hat{r}))},
\end{equation}
with
\begin{equation}
x(\hat{r})= -\left(\frac{\hat{r}_{\rm c}}{\hat{r}} \right)^{4} \, \exp\left[-  2 \,  \hat{v}^{2}_{\rm{crit}} \left( \frac{1}{\hat{r}} - \frac{1}{\hat{r}_{\rm c}}  \right)
- 2 \left(  I_{\hat{g}_{V92}^{\rm{line}}}(\hat{r}) -  I_{\hat{g}_{V92}^{\rm{line}}}(\hat{r}_{\rm c})   \right)  - 1 \right] \,,
\end{equation}
where 
\begin{align}
I_{\hat{g}_{V92}^{\rm{line}}}(\hat{r})  =& \left(10^{-\delta} \, (1+\alpha)\right)^{\frac{1}{1-\alpha}} \, \left(1+\sqrt{2} \sqrt{-\frac{\left(10^{-\delta} \, (1+\alpha)-1\right)^{\frac{1}{\alpha -1}}}{\alpha}}\right)^{\alpha} \nonumber \\
& \times  (1.76 \, \beta)^{\alpha} \, G\,M_{*} \left(\frac{\hat{r}-1}{\hat{r}}\right)^{1+\gamma_{\rm v}}   \frac{\Gamma -1}{\left( a^2 [\alpha -1] (1+\gamma_{\rm v}) \, R_{*} \right) }.
\end{align}
As was mentioned in the previous section, $\hat{r}_{\rm c}$ can be obtained numerically, making the RHS of Eq. (\ref{villata-motion}) equal zero. In order to obtain the terminal velocity in a simpler way, we can use the average value of $\hat{r}_{c}$ ($\overline{\hat{r}}_{c}= 1.0026$) obtained by \citet{araya2014}. Note that this value can be used only in the supersonic region.

Equation (\ref{sol-V-MV}) has the advantage that it is based not only on the Lambert$W$ function but also on stellar parameters and the line force parameters. For a wide range of spectral types, stellar and force multiplier parameters are given \cite[see, e.g.,][]{abbott1982,ppk1986,lamers1999,noebauer2015,gormaz2019,lattimer2021}.

\begin{figure}[h]
\center
\includegraphics[width=9cm]{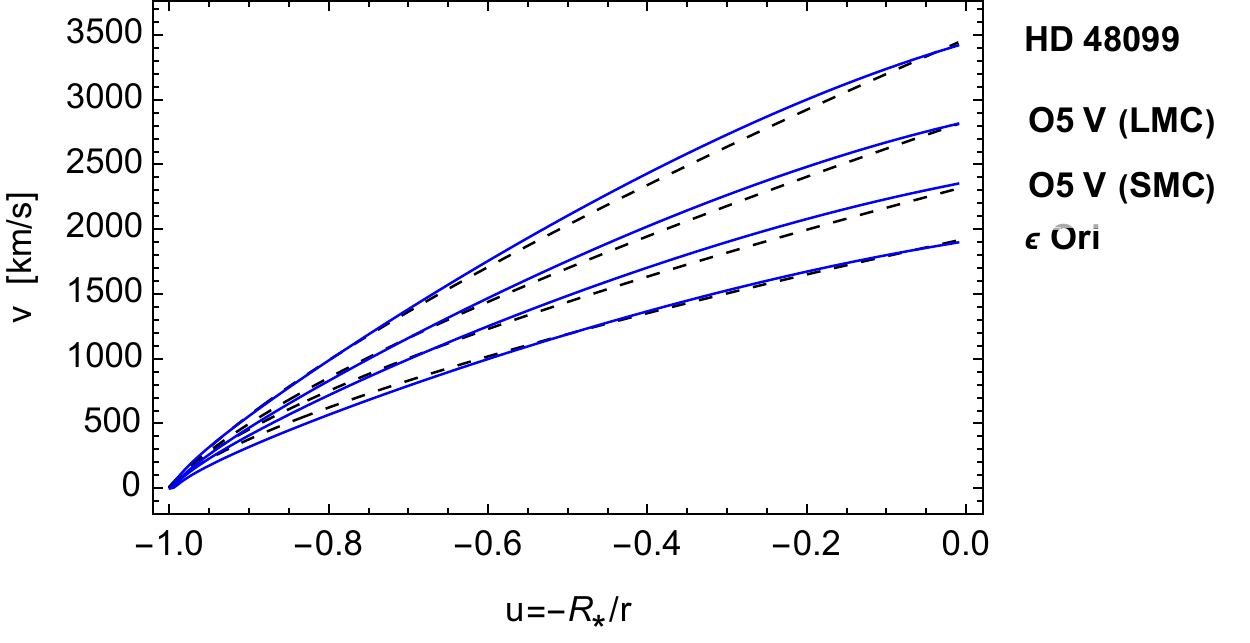}
\caption{Velocity profiles as function of the inverse radial coordinate $u=-R_{*}/r=-1/\hat{r}$ for four models. The hydrodynamic results from {\sc Hydwind} are shown in solid blue lines, and the analytical solutions are in dashed lines.  The stellar and line-force parameters for the models are given in \citet{araya2014}.  
\label{velfast}}
\end{figure}

\begin{figure}[h]
\center
\includegraphics[width=7cm]{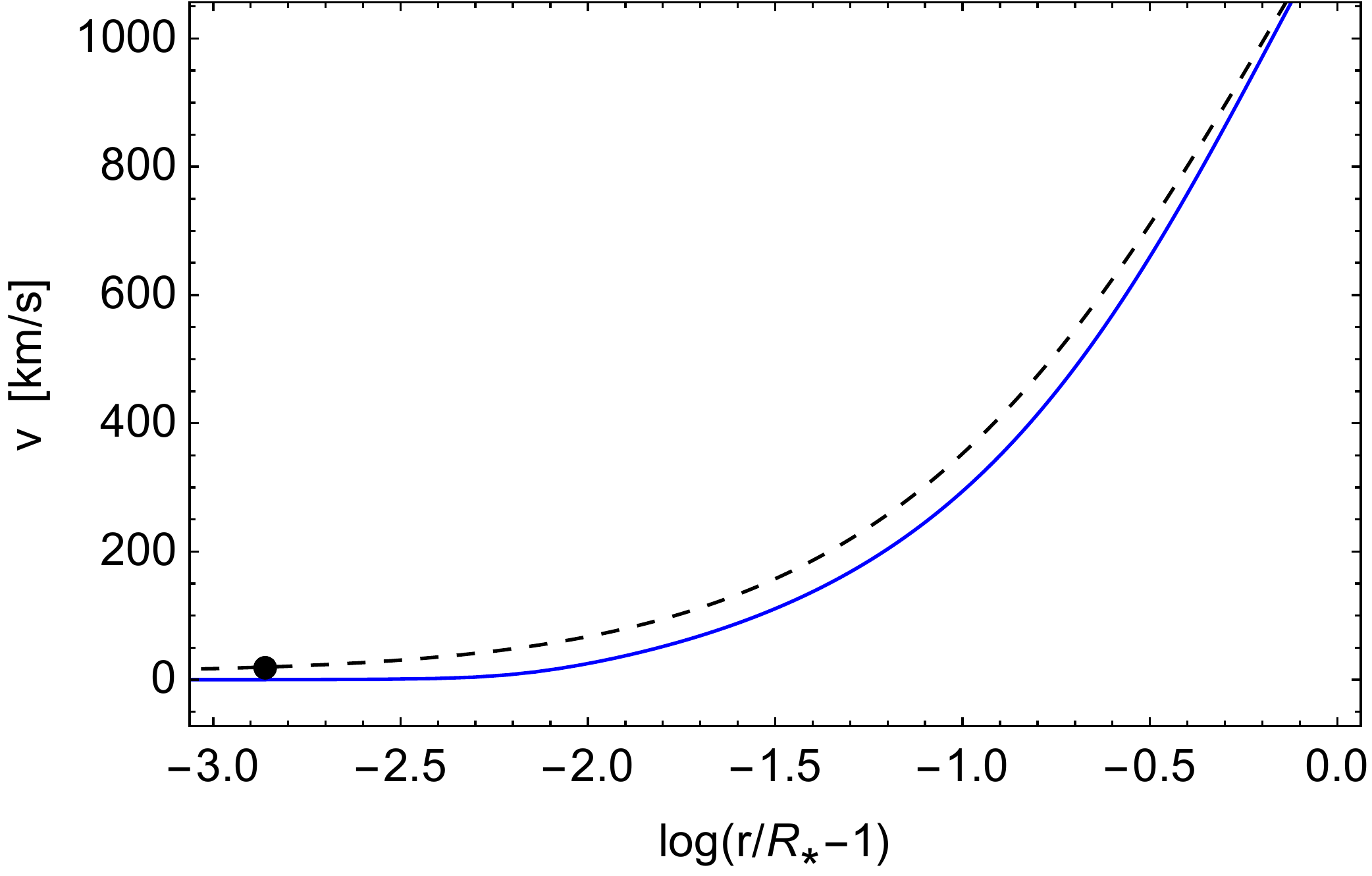}
\caption{Velocity profiles of $\epsilon$ Ori as a function of $\log(r/R_{\ast}-1)$ in a region near to the stellar surface. The solid blue line shows the numerical hydrodynamic result, and the analytical solution is in dashed line. The dot symbol indicates the position of the sonic (or critical) point. The difference between both curves is around one thermal speed.
\label{velfast-zoom}}
\end{figure}

By comparing the analytical solution to the 1-D hydrodynamic code {\sc Hydwind}, the accuracy of the analytical solution can be tested. Figure \ref{velfast} compares the results obtained with our analytical approximation to those obtained with the hydrodynamics for four stars taken from \citet{araya2014}. Both solutions have similar behaviours. However, as shown by \citet{araya2014}, the analytical approximation close to the stellar surface (subsonic region) is not good enough. In the same way, Figure \ref{velfast-zoom} compares the numerical and analytical velocity profiles near to the stellar surface for $\epsilon$ Ori. 

\begin{figure}[h]
\center
\includegraphics[width=7cm]{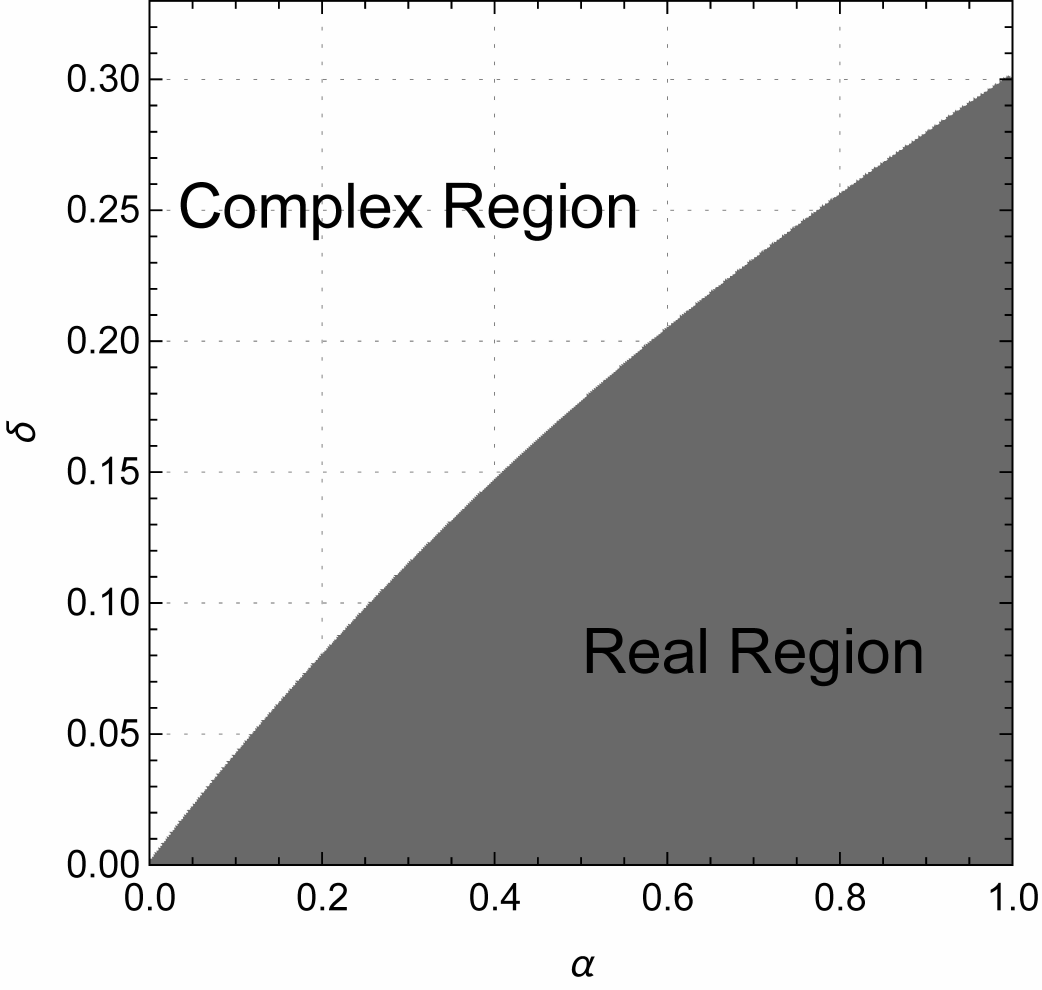}
\caption{Real and complex regions where the line acceleration expression given by \citet{villata1992} can be found. These regions are delimited by the values of the line-force parameters $\alpha$ and $\delta$.
\label{villataregion}}
\end{figure}

There is a limitation to this analytical expression when the line force parameter $\delta$ exceeds about $0.3$. This is due to the complexity of a term in the proposed line acceleration expression. To obtain an expression with real values, high values of $\delta$ would require high values of $\alpha$. However, such kind of $\alpha$ values would be totally unphysical ($\alpha > 1$). As an illustration of the dependence of this expression on the parameters $\alpha$ and $\delta$, Fig. \ref{villataregion} shows the domain of the complex and real regions when this expression is evaluated to the given line acceleration term.

\subsection{The $\delta$-slow regime}

Considering the results obtained when using an approximate description of the wind velocity for the $\delta$-slow case, \citet{araya2021} modified the function of the line acceleration given by \citetalias{muller2008} to better describe of the $\delta$-slow wind.  

As a result, the proposed line acceleration is:

\begin{equation}
\label{S-gline}
\hat{g}^{\rm{line}}(\hat{r})= \frac{\hat{g}_{0}}{\hat{r}^{1+ \delta_{1}}} \left(  1-\frac{1}{\hat{r}^{\delta_{2}}} \right) ^{\gamma},
\end{equation}
where $\hat{g}_{0}$, $\delta_{1}$, $\delta_{2}$, and $\gamma$ are the new set of line acceleration parameters. 

It is notable that the $\delta_2$ parameter, which has been incorporated into this new expression, provides a much better agreement with the numerical line acceleration obtained from the m-CAK model in the $\delta$-slow regime compared with the one from \citetalias{muller2008}. 

Based on this new definition of the radiation force, the new dimensionless equation of motion reads: 
\begin{equation}
\label{new-motion}
\left( \hat{v} - \frac{1}{\hat{v}} \right) \frac{d\hat{v}}{d \hat{r}}=
-\frac{\hat{v}_{{\rm crit}}^{2}}{\hat{r}^{2}} + \frac{2}{\hat{r}} 
+ \frac{\hat{g}_{0}}{\hat{r}^{1+ \delta_{1}}} \left(  1-\frac{1}{\hat{r}^{\delta_{2}}} \right) ^{\gamma}.
\end{equation}
The Lambert$W$ function is used to solve the equation of motion, Eq. (\ref{new-motion}) following the same methodology developed by \citetalias{muller2008},
\begin{equation}
\label{sol-S-MV}
\hat{v}(\hat{r})= \sqrt{-W_{j}(x(\hat{r}))}\, ,
\end{equation}
with
\begin{eqnarray}
\nonumber
x(\hat{r}) =  -\left(\frac{\hat{r}_{\rm c}}{\hat{r}} \right)^{4} \, \exp\left[-  2 \,  \hat{v}^{2}_{\rm{crit}} \left( \frac{1}{\hat{r}} - \frac{1}{\hat{r}_{\rm c}}  \right) - 2 \left(  I_{\hat{g}^{\rm{line}}}(\hat{r}) -  I_{\hat{g}^{\rm{line}}}(\hat{r}_{\rm c})   \right)  - 1 \right],
\end{eqnarray}

\noindent where 

\begin{eqnarray}
I_{\hat{g}^{\rm{line}}} = \int \hat{g}^{\rm{line}}(\hat{r}) d \hat{r} 
 =-\frac{g_{0}\, \hat{r}^{-\delta_{1}} \,{_{2}F_{1}} \left[-\gamma ,\frac{\delta_{1}}{\delta_{2}},1+\frac{\delta_{1}}{\delta_{2}},\hat{r}^{-\delta_{2}} \right]}{\delta_{1}},
\end{eqnarray}

\noindent 
being ${_{2}F_{1}}$ the Gauss Hypergeometric function. The critical (or sonic) point, $\hat{r}_{\rm c}$, is obtained numerically, making the RHS of Eq. (\ref{new-motion}) equal zero.\\

Ultimately, this expression for the velocity profile is in quite satisfactory agreement with the numerical solution from {\sc Hydwind}.

As described in \citet{araya2014}, a relationship was established between the \citetalias{muller2008} line-force parameters ($\hat{g_{0}}$, $\delta_{1}$, $\hat{r_{0}}$, and $\gamma$) and the stellar and m-CAK line-force parameters. In addition to being easy to use, this relationship provides a straightforward and versatile method of calculating velocity profiles analytically for a wide range of spectral types since both stellar and m-CAK line force parameters are available \citep[see,][]{abbott1982,ppk1986,lamers1999,noebauer2015,gormaz2019,lattimer2021}. 

A similar relationship can be derived for the $\delta$-slow regime using m-CAK hydrodynamic models, that is, creating a grid of {\sc Hydwind} models for $\delta$-slow solutions. These models are then analysed using a multivariate multiple regression analysis \citep[MMR][]{rencher2012,mardia1980}. 

\begin{figure}
\center
\includegraphics[width=7cm]{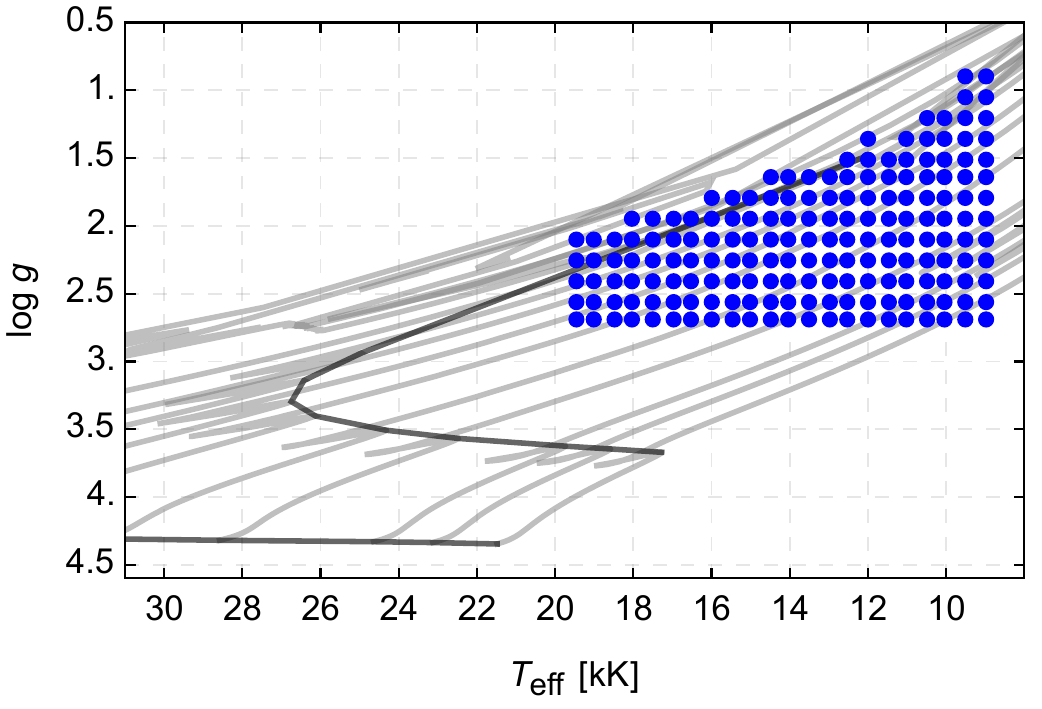}
\caption{Hydrodynamic models in the  $T_{\rm{eff}}$-$\log\,g$ plane. Blue dots represent the converged solutions. Grey solid lines are the evolutionary tracks for stars of $7 M_{\odot}$ to $60 M_{\odot}$ without rotation \citep{ekstrom2012}, and black lines represent the zero-age main sequence (ZAMS) and the terminal age main-sequence (TAMS).
\label{Dgrid}}
\end{figure}

To develop this hydrodynamic grid, the stellar radius is calculated from $M_{\rm{bol}}$ for each pair of stellar parameters ($T_{\rm{eff}}$, $\log\,g$) by using the flux weighted gravity-luminosity relationship \citep{kudritzki2003,kudritzki2008}. Additionally, a total of 20 stellar radius values were added (ranging from 5 $R_{\odot}$ to 100 $R_{\odot}$ in steps of 5 $R_{\odot}$). Surface gravities are in the range of $\log g = 2.7$ down to about $90\%$ of Eddington's limit in steps of 0.15 dex.
Effective temperatures are between $9\,000$ K to $19\,500$ K, in steps of 500 K. The range of this grid has been chosen to cover the region of the $T_{\rm{eff}}$-$\log\,g$ diagram that contains  B- and A-type supergiants. In Table \ref{Stable1}, the  m-CAK line-force parameters for each set of $(T_{\rm{eff}}$, $\log\,g$) values are listed. For the purpose of obtaining $\delta$-slow solutions, only high values of $\delta$ are considered. For the $T_{\rm{eff}}$-$\log\,g$ plane (see Fig. \ref{Dgrid}), we show in blue dots all converged models.

\begin{table}[H] 
\caption{m-CAK line-force parameter ranges for the grid of models. \label{Stable1}}
\newcolumntype{C}{>{\centering\arraybackslash}X}
\begin{tabularx}{\textwidth}{CCC}
\toprule
\textbf{Parameter}	& \textbf{Range}\\
\midrule
$\alpha$ & 0.45 - 0.69 (step size of 0.02) \\
$k$        & 0.05 - 1.00 (step size of 0.05) \\
$\delta$ & 0.26 - 0.35 (step size of 0.01)\\ 
\bottomrule
\end{tabularx}
\end{table}
\unskip

In order to obtain the new line acceleration parameters ($\hat{g}_{0}$, $\delta_{1}$, $\delta_{2}$, and $\gamma$) for each model, the m-CAK line acceleration was fitted, using Least Squares, with the proposed line acceleration expression (Eq. \ref{S-gline}).
Then, a MMR is applied to the grid of models in order to derive the relationship between new line acceleration parameters ($\hat{g}_0$, $\delta_1$, $\delta_2$ and $\gamma$) and stellar ($T_{\mathrm{eff}}$, $\log\mathrm{g}$, $R_{*}/R_{\odot}$) and m-CAK line-force parameters ($k$, $\alpha$, $\delta$). The estimated parameters are:

\begin{eqnarray}\label{mod:g_0}
\hat{g}_0^{0.27} &=& -4.548-  1.890 \times 10^{-4}  \, T_{\mathrm{eff}} +   \\
& & 4.393 \, \log\mathrm{g}+ 3.026 \times 10^{-2}  R_{*}/R_{\odot} -  \nonumber \\
& & 4.802 \times 10^{-3} \,  k + 3.781   \, \alpha - 3.212   \, \delta, \nonumber
\end{eqnarray}

\begin{eqnarray}\label{mod:delta1}
(\delta_1+1)^{5.3} &=& -4.623-  3.743 \times 10^{-4}  \, T_{\mathrm{eff}} +   \\
& & 1.489 \times 10^{1} \, \log\mathrm{g}+ 1.148 \times 10^{-1}  R_{*}/R_{\odot} +  \nonumber \\
& & 2.415 \,  k + 9.553 \times 10^{1}   \, \alpha - 1.320 \times 10^{2}   \, \delta, \nonumber
\end{eqnarray}

\begin{eqnarray}\label{mod:delta2}
\delta_2^{0.45} &=& 5.359 +  8.262 \times 10^{-5}  \, T_{\mathrm{eff}} -   \\
& & 1.327 \, \log\mathrm{g} - 8.327 \times 10^{-3}  R_{*}/R_{\odot} +  \nonumber \\
& & 2.181 \times 10^{-1} \,  k + 9.618 \times 10^{-1}  \, \alpha - 2.296   \, \delta \nonumber
\end{eqnarray}

\noindent and

\begin{eqnarray}\label{mod:gamma}
(\gamma + 1)^{-3.56} &=& -1.031 +  7.254 \times 10^{-6}  \, T_{\mathrm{eff}} +   \\
& & 2.994 \times 10^{-1} \, \log\mathrm{g}+ 3.097 \times 10^{-3}  R_{*}/R_{\odot} +  \nonumber \\
& & 1.836 \times 10^{-1} \,  k - 4.828 \times 10^{-1}  \, \alpha + 1.254   \, \delta, \nonumber 
\end{eqnarray}

After the estimated values for each dependent variable, $\hat{g}_0^{0.27}$, $(\delta_1+1)^{5.3}$, $\delta_2^{0.45}$, $(\gamma + 1)^{-3.56}$,  are obtained  they are transformed into $\hat{g}_0$, $\delta_1$, $\delta_2$, and  $\gamma$ through their respective inverse functions. Finally, we can use these parameters in Eq. (\ref{sol-S-MV}) to calculate the velocity profile.

\begin{figure}[h]
\center
\includegraphics[width=9cm]{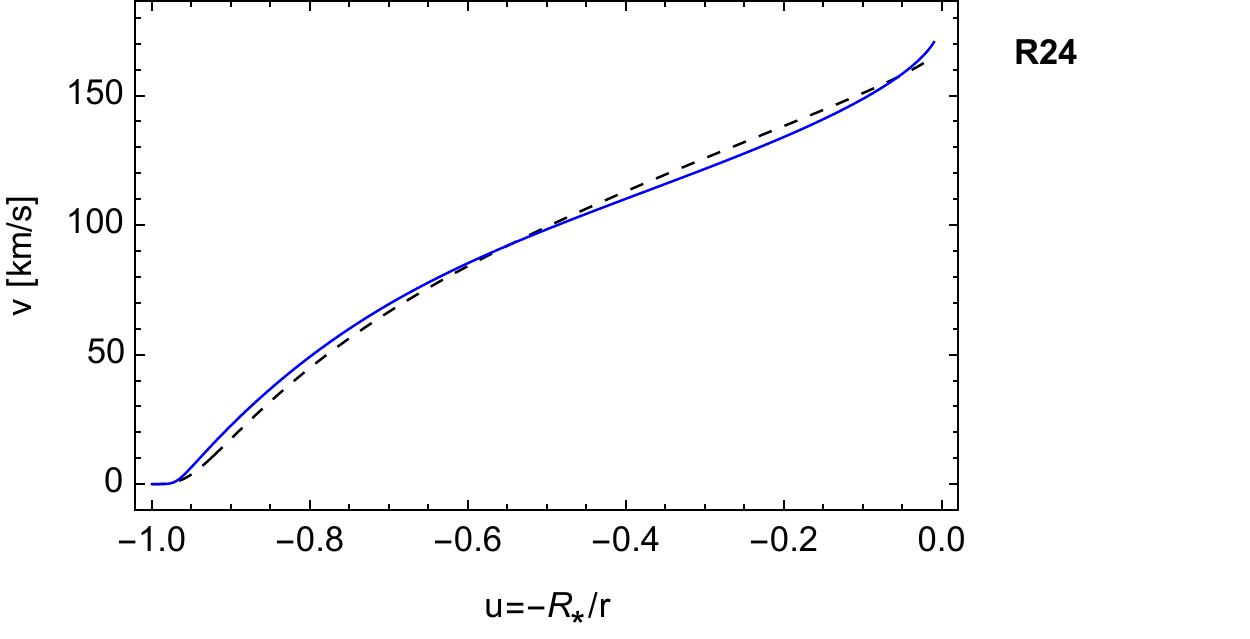}
\caption{Velocity profiles as function of the inverse radial coordinate $u=-R_{*}/r=-1/\hat{r}$ for the model R24 from \citet{cure2011}. The hydrodynamic result from {\sc Hydwind} is shown in the solid blue line, and the analytical solution is a dashed black line. 
\label{velslow}}
\end{figure}

The velocity profiles obtained via {\sc Hydwind} code and the analytical solution are shown in Fig. \ref{velslow} for one model with $\delta$-slow solution. The model is taken from \citet{cure2011}.

Remember, however, that this relationship holds only for $\delta$-slow solutions, especially for values of $\delta$ between $0.29$ and $0.35$. In addition, considering the number of converged models in the grid, the authors recommend using this expression for values of $\alpha$ between $0.45$ and $0.55$.

Finally, it is important to remark that both analytical solutions for the velocity profile, fast and $\delta$-slow, do depend only on the stellar ($T_{\mathrm{eff}}$, $\log\mathrm{g}$, $R_{*}/R_{\odot}$) and m-CAK line-force parameters ($k$, $\alpha$, $\delta$). Regarding to the mass-loss rate, \citetalias{villata1992} proposed an expression to obtain it (in terms of the stellar and m-CAK parameters) following the approximations given by \citet{kppa1989}. Also, \citet{araya2021}, in the appendix, proposed a method to obtain the mass-loss based on \citet{cure2004}.

\section{Summary and discussion \label{DiscConclu}}

Observations over the past decades have shown that the basic wind parameters behave as predicted by the theory. This fundamental agreement between observations and theory provides strong evidence that the winds from massive stars are driven by radiation pressure. This has given the m-CAK theory a well-established status in the massive star community. However, several issues are contentious and still unclear, such as the calibration of the Wind-momentum Luminosity Relationship (WLR) \citep{kudritzki1995}, disks of Be stars, wind parameters determination, the applicability of the called slow wind solutions, among others. All these issues are the focus of massive stars research. 

This review is focused on the theoretical and numerical research of wind hydrodynamics of massive stars based on the m-CAK theory, with particular emphasis on its topology and hydrodynamic solutions. 

We presented a topological analysis of the one-dimensional m-CAK hydrodynamic model and its three known hydrodynamic solutions, the fast, $\Omega$-slow and $\delta$-slow solutions. From a topological point of view, slow solutions are obtained from a new branch of solutions with a locus of singular points far from the stellar surface, unlike fast solutions with a family of singular points near the stellar surface.

We continued analyzing the dependence of the line force parameters ($k$, $\alpha$, and $\delta$) with the wind parameters (mass loss rate and terminal velocity) in order to understand the complex non-linear dependence between these parameters. In the case of $\alpha$, there is an increase in both wind parameters as this parameter increases. This behaviour is similar to the $k$ parameter, but the dependence is very slight for terminal velocity, while the mass loss rate has a significant impact. For the $\delta$ parameter, the terminal velocity has a decreasing behaviour when this parameter increases, while the mass-loss rate can have a decreasing or increasing behaviour, which depends on the parameter $k$. When $k$ is low, mass-loss rates decrease, while $\delta$ increases, whereas when $k$ is large, the opposite occurs.

In addition, we compared the $\beta$-law with the hydrodynamic solutions. We concluded that the fast solution could not be adequately described by a $\beta$-law with $\beta>1.2$, while the $\delta$-slow solution cannot be described by any $\beta$-law.

Furthermore, we presented two analytical expressions for the solution of line-driven winds in terms of the stellar and line-force parameters. The expressions are addressed to obtain the fast and $\delta$-slow velocity regimes in a simple way. Both solutions are based on the Lambert$W$ function and an approximative expression for the wind line acceleration as a function of the radial distance. The importance of an analytical solution lies in its simplicity in studying the properties of the wind instead of solving complex hydrodynamic equations. In addition, these analytical expressions can be used in radiative transfer or stellar evolution codes \citep[see, e.g.,][]{gormaz2022}. 

Concerning the applicability of the slows solutions, in the case of the $\Omega$-slow solution, their behaviour suggest that it can play a paramount role in the ejection of material to the equatorial circumstellar environment of Be and B[e] stars. 
Be stars are a unique set of massive stars whose main distinguishing characteristics are rapid rotation and the presence of a dense, gaseous circumstellar disk orbiting in a quasi-Keplerian fashion. There is a long-standing problem in understanding the formation of disks in Be stars; this is one of the major areas of ongoing research in Be stars. The gaseous disks are not remnants of the objects’ protostellar environments, nor are they formed through the accretion of material \cite{porter2003, rivinius2013}. On the contrary, the equatorial gas consists of a decretion disk formed from a material originating from the central star. 

As was stated above, attempts to solve this problem have been made without much success, for example, the link between the line-driven winds and these discs, called the Wind Compressed Disk \citep{bjorkman1993}, but the work of \citet{owocki1996} was the first to show this is not a viable mechanism for rapidly rotating stars due to the non-radial line force components. The most accepted model to successfully reproduce many Be star observables is the viscous decretion disk (VDD) model, developed by \citet{lee1991} and examined by \citet{okazaki2001}, \citet{bjorkman2005}, \citet{krticka2011} and \citet{cure2022}. Currently, how that material is ejected into the equatorial plane and how sufficient angular momentum is transferred to the disk to maintain quasi-Keplerian rotation are among the primary unresolved questions currently driving classical Be star research.

\citet{araya2017} studied the $\Omega$-slow wind solution and its relation with the disks of Be stars. Overall, this work is an extension to the study done by \citet{silaj2014b}, where they precisely investigated if the density distribution provided by the $\Omega$-slow wind solution could adequately describe the physical conditions to form a dense disc in Keplerian rotation via angular momentum transfer. They considered a two-component wind model, i.e.,  a fast, thin wind in the polar latitudes and a $\Omega$-slow, dense wind in the equatorial regions. Based on the equatorial density distributions, H$\alpha$ line profiles are generated and compared with an ad-hoc emission profile, which agrees with the observations. In addition, their calculations assumed three different scenarios related to the shape (oblate correction factor) and the star's brightness (gravity darkening). Finally, they found that under certain conditions (related to the line-force parameter of the wind), a significant H$\alpha$ line profile could be produced, and maybe the line-driven winds through the $\Omega$-slow solution can have an essential role in the disc formation of Be stars.

In addition, \citet{araya2018} studied the zone where the classical m-CAK fast solution 
ceases to exist, and the $\Omega$-slow solution emerges at rapid rotational speeds. This study used two hydrodynamic codes with time-dependent and stationary approaches.  They found that both solutions can co-exist in this transition region, which depends exclusively on the initial conditions. In addition, they performed base-density wind perturbations to test the stability of the solution within the co-existence region. A switch of the solution was found under certain perturbation conditions. The results are addressed to a possible role in the ejection of extra material into the equatorial plane via pulsation modes, where the $\Omega$-slow solution can play an important role.  

A current weakness of this m-CAK model is that it does not consider the role of viscosity and its influence on angular momentum transport. This mechanism might explain the formation of a Keplerian disc.\\

On the other hand, the  $\delta$-slow solution is promising to explain the discrepancies of the wind parameters between observations and theory in late B- and A-type supergiant stars. 
According to the findings of \citet{venero2016}, these suggest that the terminal velocity of early and mid-B supergiants agrees with the results seen from fast outflowing winds. By contrast, the results obtained for late B supergiants and, mainly, those obtained for early A supergiants agree with the results achieved for $\delta$-slow stationary outflowing wind regimes. Then, the $\delta$-slow solution enables us to describe the global features of the wind quite well, such as mass-loss rates and terminal velocities of moderately and slowly rotating B supergiants. 

Conversely, \citet{venero2016} stated that the $\delta$-slow solution seems not to be present in stars with $T_{\rm{eff}}>21 \,000$ K. This restricts the possibility of a switch between fast and slow regimes at such temperatures. Consequently, this would be a physical explanation for why an empirical bi-stability jump can be observed around $21\, 000$ K in B supergiants \citep{lamers1995}. From a theoretical perspective, a velocity jump has also been found using Monte Carlo modelling and the co-moving frame method \cite[see, e.g.,][]{vink2018,vink2021,vink2022}

In addition, it is generally accepted that most O and early B-type stars can be adequately modelled with a $\beta$ velocity law with $0.7 \lesssim \beta \lesssim 1.2$. However, supergiants A and B exhibit $\beta$ values that tend towards higher values, often $\beta \ge 2$ \citep[see, e.g.,][]{stahl1991,lefever2007,markova2008,haucke2018, rivet2020}. Venero et al. 2023 (in preparation) propose that $\delta$-slow solutions might explain these winds.  They show that the $\delta$-slow regime could adequately fit the H$\alpha$ line profile of B supergiants with the same accuracy as that obtained using a $\beta$-law with $\beta \ge 2$, but now with a hydrodynamic explanation of the velocity profile used.\\

The investigation carried out in the latest works inspired us to go deeper into the possible role of slow wind solutions with respect to the  unresolved questions related to massive stars. In view of our results, we are encouraged to develop this line of research further.  In the case of the $\Omega$-slow solution and its link to Be stars, or possibly to B[e] stars, it would require 2D/3D models for a better understanding to take  into account non-radial forces, the effects of stellar distortion and gravity darkening. These considerations could change, in turn, the nature of the $\Omega$-slow solution or the behaviour regarding the co-existence of solutions and a switch between them. \\
The $\delta$-slow solution could play an essential role in understanding the winds of B- and A-type supergiants. Moreover, this solution is expected to solve the disagreement between the observations and theory for these stars and, in this way, calibrate the wind-momentum luminosity relationship. 

As we mentioned previously. In the standard procedure for finding stellar and wind parameters, the $\beta$-law ($\beta$, $v_{\infty})$ and a mass-loss rate ($\dot{M}$) are three \textit{'free'} input parameters in radiative transfer codes, comparing synthetic spectra with the observed spectra of a star. The $\beta$ law comes from an approximation of the fast  wind solutions, and the values of $\beta$ should be in a restricted interval. To improve the hydrodynamic approximation used in this standard procedure, we have developed two hydrodynamic procedures to  derive stellar and wind parameters: 
\begin{itemize}
    \item The self-consistent CAK procedure \citep{gormaz2019}, based on the m-CAK model. Here we iteratively calculate the line-force parameters using the atomic line database from CMFGEN, coupled with the m-CAK hydrodynamic until convergence. We obtain the line-force parameters and, therefore, the velocity profile and the mass-loss rate. Thus, none of the input parameters are \textit{'free'}, but self-consistently calculated.
    \item The Lambert$W$ procedure \citep{gormaz2021}. In this procedure, we start using a $\beta$-law and a value for $\dot{M}$ in CMFGEN. After convergence, we calculated the line acceleration as a function of $r$, and using the Lambert$W$ function we obtain a new velocity profile. This is inserted in CMFGEN and the cycle is repeated until convergence. In this Lambert$W$ procedure, the only input parameter is the mass-loss rate.
\end{itemize}
We expect that these two alternatives, which reduce the number of input parameters, will in the future, have a significant impact on the standard procedures for obtaining stellar and wind parameters of massive stars.

\authorcontributions{All authors have read and agreed to the published version of the manuscript.}

\funding{We are grateful for the support from FONDECYT projects 1190485 and 1230131. I.A. also thanks for the support from FONDECYT project 11190147. This project has received funding from the European Union’s Framework Programme for Research and Innovation Horizon 2020 (2014-2020) under the Marie Sk\l{}odowska-Curie Grant Agreement No. 823734.}

\acknowledgments{We are grateful for suggestions and comments from reviewers to improve this work. We would like to thank the continuous support from Centro de Astrofísica de Valparaíso and our colleagues and students from the massive stars group at Universidad de Valpara\'iso, especially to Catalina Arcos. We also thank our long-standing colleagues from other institutes who have contributed to our understanding of the field, especially to Lydia Cidale, Diego Rial and Roberto Venero. We also wish to acknowledge the support received from our respective Universities, Universidad de Valpara\'iso and Universidad Mayor, to continue our research.}

\conflictsofinterest{The authors declare no conflict of interest.}

\abbreviations{Abbreviations}{
The following abbreviations are used in this manuscript:\\

\noindent 
\begin{tabular}{@{}ll}
ZAMS & Zero-Age Main Sequence \\
EoM & Equation of Motion\\
MV08 & \citet{muller2008}\\
KPPA & \citet{kppa1989}\\
KPP & \citet{kpp1987}\\
CAK & \citet{castor1975}\\
FA & \citet{friend1986}\\
RHS & Right Hand Side\\
MMR & Multivariate Multiple Regression\\
WLR & Wind-momentum Luminosity Relationship \\
VDD & Viscous Decretion Disk
\end{tabular}
}

\appendixtitles{no} 
\appendixstart

\appendix
\section[\appendixname~\thesection ]{\label{appA}}

In their original work, \citetalias{castor1975} discussed the stellar point approximation of their model and properly estimated the influence of the disk correction factor ($CF$) on the wind dynamics, i.e.,
reducing the line force in about 40\% at the stellar surface.\\

The definition of the $CF$ is: 
\begin{equation} 
CF=\frac{2}{1-\mu _{\ast}^{2}}\int_{\mu _{\ast}}^{1} \left( \frac{ 
(1-\mu ^{2})v/r+\mu ^{2}v^{' }}{v^{' }}\right) ^{\alpha }\mu d\mu . 
\label{eqa1} 
\end{equation}
where $v'=dv/dr$ and $\mu _{\ast}^2={1-(R_{\ast}/r)^{2}}$.\\
Integrating \ref{eqa1} and changing the variables from $r\,\xrightarrow{}\,u=-R_{\ast }/r$, and $v\,\xrightarrow{}\,w=v/a$, where $a\;$is the
thermal speed); the finite disc correction factor transforms to: 
\begin{equation} 
CF(u,w/w')=\frac{1}{1-\alpha
}\frac{1}{u^{2}}\frac{1}{(1+\frac{1}{u}\frac{w}{w^{' }})}\left[
1-\left( 1-u^{2}-u\frac{w}{w'}\right)^{(1+\alpha )}\right]\, , 
\label{eqa2} 
\end{equation} where $w^{' }=dw/dr$. Due to the fact that $CF$
depends on $u$ and the ratio $Z = w/w'$, we can define
$\lambda$ as:
\begin{equation}
\lambda = (u+Z)\; u .
\label{eqa2b} 
\end{equation} Therefore, re-writing $CF(u,Z)$ as: 
\begin{equation} 
CF(\lambda)=\frac{1}{(1-\alpha)}\frac{1}{\lambda} \left[1-\left(1-\lambda
\right)^{(1+\alpha)} \right].  \label{eqa3} 
\end{equation} 
 
The partial derivatives of $CF$ with respect to $u,w,w^{' }$ are then
related to $\partial CF/\partial \lambda $ via the chain rule, namely:
\begin{eqnarray}
\frac{\partial CF}{\partial u}&=&\frac{\partial CF}{\partial \lambda }\times \frac{ 
\partial \lambda }{\partial u}.  \label{eqa4} \\
\frac{\partial CF}{\partial w}&=&\frac{\partial CF}{\partial \lambda }\times \frac{ 
\partial \lambda }{\partial w}.  \label{eqa5} \\ 
\frac{\partial CF}{\partial w^{' }}&=&\frac{\partial CF}{\partial \lambda  
} \times \frac{\partial \lambda }{\partial w^{' }}.  \label{eqa6}
\end{eqnarray}  
The function  $e(\lambda )=\partial CF/\partial \lambda $ is therefore:
\begin{equation} 
e(\lambda )=\frac{ \left( 1- \lambda \right)^{\alpha} - CF(\lambda)}{\lambda}. 
\label{eqa7} 
\end{equation} 
then (\ref{eqa4}), (\ref{eqa5}) and (\ref{eqa6}) are related to (\ref{eqa7}) 
by:  
\begin{equation} 
e(\lambda )=\frac{1}{2u+Z} \times \frac{\partial CF}{\partial
u}=\frac{w'}{u} \times \frac{\partial CF}{\partial w}=-\frac{w^{'
}}{uZ} \times \frac{\partial CF}{ 
\partial w^{' }}.  \label{eqa8} 
\end{equation} 
Approximating $Z$ by a $\beta$-field ($Z=(1+u)/\beta$), 
we obtain $CF$ and $e$ {\it only }as functions of $u$. 


\section[\appendixname~\thesection ]{\label{appB}}
Here the basic steps toward equations (\ref{2.9a}), (\ref{2.9b}) and (\ref 
{2.9c}) are outlined. The reader should keep in mind the original derivation by \citetalias{castor1975}.
 
The partial derivatives of $F(u,w,w')$ (eq. \ref{2.9a}), with respect 
to $u,w$ and $w'$ are:  
\begin{equation} 
\frac{\partial F}{\partial u}= -\frac{2}{u^{2}} + a^{2}_{rot} - C^{' } 
\left( \frac{\partial CF}{\partial u}g + CF\frac{\partial g}{\partial u} 
\right) \;w^{-\delta }\;(w\,w')^{\alpha }\;  \label{a1} 
\end{equation} 
\begin{equation} 
\frac{\partial F}{\partial w}=\left( 1+\frac{1}{w^{2}}\right) w^{' }- 
C^{' }\left( \frac{\partial CF}{\partial w}+\alpha \frac{CF}{w} 
-\delta \frac{CF}{w}\right) \;g\;w^{-\delta }\;(w\,w')^{\alpha }  
\label{a2} 
\end{equation} 
\begin{equation} 
\frac{\partial F}{\partial w'}=\left( 1-\frac{1}{w^{2}}\right) w 
-C^{' }\left( \frac{\partial CF}{\partial w'}+\alpha 
\frac{CF}{w'}\right) \;g\;w^{-\delta }\;(w\,w')^{\alpha }   
\label{a3} 
\end{equation} 
 
After using the new coordinate $Y=w\;w^{' }$ and $Z=w/w^{' }$, 
some derivative relation of the correction factor (see appendix \ref{appA}) and defining:  
\begin{equation} 
\frac{dg(u)}{du}=g(u) \times h(u)  \label{a4} 
\end{equation} 
where  
\begin{equation} 
h(u)=\delta \left( \frac{2}{u}-\frac{u}{\sqrt{1-u^{2}}\left(
1-\sqrt{1-u^{2}}\right) }\right)  \label{a5} 
\end{equation} 
the singularity condition ($w' \;\partial F/\partial w'=0$) now
reads:  
\begin{equation} 
\left(1-\frac{1}{YZ}\right)\;Y\;-C' 
\;f_{2}(u,Z)\;g\;Z^{-\delta /2}\;Y^{\alpha -\delta /2}=0  \label{a6} 
\end{equation} 
here$\;f_{1}(u,Z) = CF(u,Z)$,  and 
\begin{equation} 
f_{2}(u,Z)=\alpha \;f_{1}(u,Z) - u\;Z\times e(u,Z)  \label{a8} 
\end{equation} 
and $e(u,Z) = e(\lambda)$ is defined in appendix \ref{appA}. 
 
The regularity condition ($Z\;dF/du=0$) now transform to:  
\begin{equation} 
\left(1+\frac{1}{YZ}\right)\;Y-C'
\;f_{3}(u,Z)\;g(u)\;Z^{-\delta /2}\;Y^{\alpha -\delta
/2}=+\frac{2Z}{u^{2}}-a_{rot}^{2}Z\label{a9} 
\end{equation} 
where 
\begin{equation} 
f_{3}(u,Z)=(3u+Z)\;Z\times e(u,Z)+f_{1}(u,Z)\times (h(u)\;Z+\alpha
-\delta )  \label{a10} 
\end{equation} 


\begin{adjustwidth}{-\extralength}{0cm}

\reftitle{References}
\bibliography{cites}

\end{adjustwidth}
\end{document}